\pdfoutput=1
\documentclass[sigconf,natbib=true,anonymous=false]{acmart}

\setcopyright{none}
\renewcommand\footnotetextcopyrightpermission[1]{}

\usepackage{xspace}
\usepackage{xcolor,colortbl}
\usepackage{mdframed}
\usepackage{arydshln}

\newcommand{\ragnarok}[0]{Ragnar\"ok\xspace}

\begin{document}

%%
%% The "title" command has an optional parameter,
%% allowing the author to define a "short title" to be used in page headers.
\title{\ragnarok: A Reusable RAG Framework and Baselines for \\ TREC 2024 Retrieval-Augmented Generation Track}

%%
%% The "author" command and its associated commands are used to define
%% the authors and their affiliations.
%% Of note is the shared affiliation of the first two authors, and the
%% "authornote" and "authornotemark" commands
%% used to denote shared contribution to the research.
\author{Ronak Pradeep}
\authornote{Both authors contributed equally to the paper. Correspondence to Ronak Pradeep <\url{rpradeep@uwaterloo.ca}> and Nandan Thakur <\url{nandan.thakur@uwaterloo.ca}>.}
\affiliation{University of Waterloo \\ \city{Waterloo}\country{Canada}}

\author{Nandan Thakur}
\authornotemark[1]
\affiliation{University of Waterloo \\ \city{Waterloo}\country{Canada}}

\author{Sahel Sharifymoghaddam}
\affiliation{University of Waterloo \\ \city{Waterloo}\country{Canada}}

\author{Eric Zhang}
\affiliation{University of Waterloo \\ \city{Waterloo}\country{Canada}}

\author{Ryan Nguyen}
\affiliation{University of Waterloo \\ \city{Waterloo}\country{Canada}}

\author{Daniel Campos}
\affiliation{Snowflake Inc. \\ \city{New York}\country{USA}}

\author{Nick Craswell}
\affiliation{Microsoft \\ \city{Seattle}\country{USA}}

\author{Jimmy Lin}
\affiliation{University of Waterloo \\ \city{Waterloo}\country{Canada}}

\settopmatter{authorsperrow=4}

%%
%% By default, the full list of authors will be used in the page
%% headers. Often, this list is too long, and will overlap
%% other information printed in the page headers. This command allows
%% the author to define a more concise list
%% of authors' names for this purpose.
\renewcommand{\shortauthors}{Pradeep and Thakur et al.}

%%
%% The abstract is a short summary of the work to be presented in the
%% article.
\begin{abstract}
Did you try out the new Bing Search?
Or maybe you fiddled around with Google AI~Overviews? 
These might sound familiar because the modern-day search stack has recently evolved to include retrieval-augmented generation (RAG) systems. 
They allow searching and incorporating real-time data into large language models (LLMs) to provide a well-informed, attributed, concise summary in contrast to the traditional search paradigm that relies on displaying a ranked list of documents.
Therefore, given these recent advancements, it is crucial to have an arena to build, test, visualize, and systematically evaluate RAG-based search systems. 
With this in mind, we propose the TREC 2024 RAG Track to foster innovation in evaluating RAG systems.
In our work, we lay out the steps we've made towards making this track a reality --- we describe the details of our reusable framework, \ragnarok, explain the curation of the new MS MARCO V2.1 collection choice, release the development topics for the track, and standardize the I/O definitions which assist the end user.
Next, using \ragnarok, we identify and provide key industrial baselines such as OpenAI's GPT-4o or Cohere's Command R+.
Further, we introduce a web-based user interface for an interactive arena allowing benchmarking pairwise RAG systems by crowdsourcing.
We open-source our \ragnarok framework and baselines to achieve a unified standard for future RAG systems. 

\vspace{0.5em}
\hspace{3.6em}\includegraphics[width=1.25em,height=1.25em]{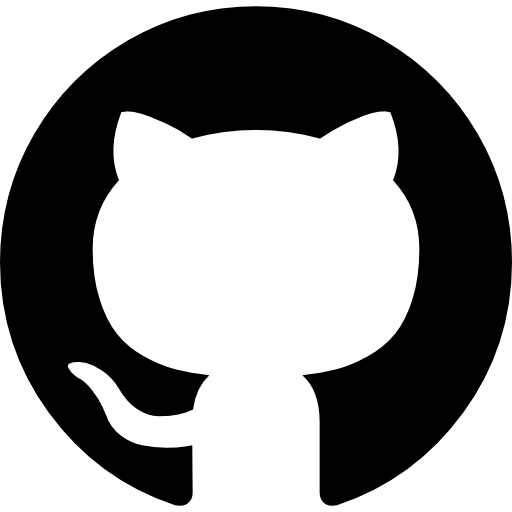}\hspace{.3em}
\parbox[c]{\columnwidth}
{
    \vspace{-.55em}
    \href{https://github.com/castorini/ragnarok}{\nolinkurl{https://github.com/castorini/ragnarok}}
}
\vspace{-1.4em}
\end{abstract}

%%
%% The code below is generated by the tool at http://dl.acm.org/ccs.cfm.
%% Please copy and paste the code instead of the example below.
%%
% \begin{CCSXML}
% <ccs2012>
% <concept>
% <concept_id>10002951.10003317.10003359.10003360</concept_id>
% <concept_desc>Information systems~Test collections</concept_desc>
% <concept_significance>500</concept_significance>
% </concept>
% <concept>
% <concept_id>10002951.10003317.10003371</concept_id>
% <concept_desc>Information systems~Specialized information retrieval</concept_desc>
% <concept_significance>500</concept_significance>
% </concept>
% </ccs2012>
% \end{CCSXML}

% \ccsdesc[500]{Information systems~Specialized information retrieval}
% \ccsdesc[500]{Information systems~Test collections}

%%
%% Keywords. The author(s) should pick words that accurately describe
%% the work being presented. Separate the keywords with commas.
% \keywords{Retrieval-Augmented Generation; RAG; Framework; Baselines}

% \received{20 February 2007}
% \received[revised]{12 March 2009}
% \received[accepted]{5 June 2009}

%%
%% This command processes the author and affiliation and title
%% information and builds the first part of the formatted document.
\maketitle

\section{Introduction}

Retrieval Augmented Generation (RAG)~\cite{guu:2020, lewis:2020, izacard:2021, borgeaud:2022} has emerged as a popular technique to augment large language model (LLM) generation for knowledge-intensive tasks such as open-domain question answering or fact verification \cite{petroni:2021}.
Using the top-$k$ retrieved segments from a suitable retrieval system, RAG systems output an answer summary grounded on the relevant context.
RAG systems mitigate factual inconsistencies in LLM outputs \cite{khandelwal:2020, lewis:2020, gao:2023, liu:2024}, and enhance interpretability~\cite{guu:2020} and generalization \cite{gao:2023b}, thus facilitating a wider adoption across several domains like Medicine~\cite{xiong:2024} and Finance~\cite{yepes:2024}.

\begin{figure*}[t]
    \centering
    \includegraphics[trim=0 5 0 10, clip, width=\textwidth]{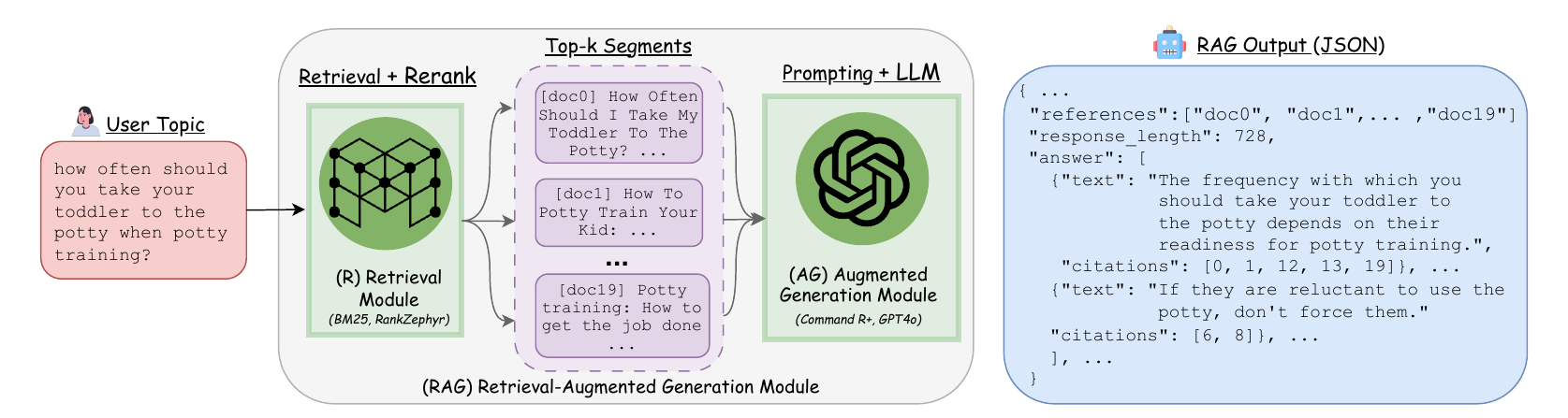}%
    \vspace{-1em}
    \caption{%
       Schematic diagram of the \ragnarok framework. 
       Given a user topic (left), the process consists of two steps: (1) 
       (R) retrieval (+ rerank), where the topic yields the top-$k$ relevant segments from our document collection (e.g., potty training articles); and (2) (AG) augmented-generation, where the retrieved segments with a suitable prompt template is fed to the large language model (LLM) to generate the post-processed answer response (JSON) containing individual sentence-level citations.
       % Finally, the answer is post-processed into a JSON containing sentence-level citations on retrieved segments as references.
       }%
    \vspace{-0.5em}
    \label{fig:ragnarok-framework}%}%
\end{figure*}

Several companies provide end-to-end RAG frameworks such as Bing Search~\cite{bingsearch}, or Google Gemini~\cite{anil:2023}. 
Most of these systems are either proprietary or offer limited user customization.
Likewise, the absence of a standardized RAG framework makes implementing RAG at a large scale challenging.
Implementing atop existing frameworks requires custom code for multiple steps including retrieval, reranking, and generation.
To promote wider adoption of RAG in academia, we develop \ragnarok, a user-friendly, reusable, end-to-end RAG framework offering code for customizable retrievers, rerankers, and generation models. 

\ragnarok comprises two key modules: (R) Retrieval and (AG) Augmented Generation. 
The retrieval module incorporates both the retrieval and re-ranking stages to yield the top-$k$ retrieved segments for an input user topic.
Next, the augmented generation module uses the user-provided topic and retrieved segments to produce an RAG answer, formatted into individual sentences, citing the relevant information from the top-$k$ retrieved segments. 
\ragnarok is deeply integrated with existing Python frameworks, such as~\texttt{Pyserini}~\cite{pyserini} and~\texttt{rank\_llm}~\cite{pradeep:2023, pradeep:2023b} and can be easily installed via PyPI using~``\texttt{pip install pyragnarok}''. 
The framework offers easy-to-use REST APIs and an integrated WebUI to enhance user-friendliness and improve the human evaluation experience.

The \ragnarok framework will be used for providing baselines in the upcoming TREC 2024 Retrieval Augmented Generation (RAG) Track.\footnote{TREC 2024 Retrieval Augmented Generation (RAG) Track: \url{https://trec-rag.github.io}.} An ideal framework should include a sufficiently large document collection covering diverse information and non-factoid, decompositional topics requiring long-form answers. 
In our framework, we deduplicate the existing MS MARCO V2 document collection. In addition, we provide a ``segment'' collection using a sliding-window chunking technique (discussed in Section \ref{sec:document-collection}). Further, we release two sets of development topics: (i) TREC-RAGgy 2024: a filtered subset of topics with long-form answers from TREC Deep Learning 2021-23 \cite{craswell:2021,craswell:2022,craswell:2024}; and (ii) TREC-Researchy 2024: a subset of the Researchy Questions introduced in \citet{rosset:2024}. 

Our \ragnarok framework supports a head-to-head RAG battle arena for the answer evaluation, heavily inspired by recent work such as the Chatbot Arena~\cite{chiang:2024, zheng:2024}.
We include key industrial baselines such as Cohere Command R+ \cite{command_r_plus} and OpenAI GPT-4o~\cite{openai_gpt4o} and evaluate both the baselines using the retrieval setup involving BM25~\cite{robertson:2009} and RankZephyr~\cite{pradeep:2023b} with human preferences. Overall, we observe GPT-4o to provide more detailed answers over Command R+ on the development set of topics (discussed in Section \ref{sec:baselines}).
Finally, we open-source \ragnarok and make it publicly available at the following URL: \url{https://github.com/castorini/ragnarok}. In the future, we will include a wider variety of LLMs as baselines and continue to improve our framework.

\section{Related Work}

\paragraph{RAG Frameworks.} 
Existing RAG systems are primarily closed-source and difficult to reproduce.
Open-source frameworks such as LangChain \cite{langchain} and LlamaIndex \cite{llamaindex}, while available, are not research-friendly and lack proper evaluation and benchmarking. 
FlashRAG~\cite{jin:2024}, a concurrent work, is a similarly motivated toolkit to improve the RAG experience for researchers. 
While the framework is extensive and designed for pipeline flexibility, \ragnarok offers a few additional capabilities --- a WebUI serving a RAG battle arena, easy-to-use REST APIs, a standardized I/O definition working with sentence-level citations, and a tight integration with popular retrieval (+ reranking) frameworks like Pyserini~\cite{pyserini} and RankLLM.

\paragraph{Collection selection.} Current RAG datasets are constructed using the English Wikipedia as the document collection, However, their scale is limited to provide rich and comprehensive information to support RAG systems. 
In addition, ClueWeb22~\cite{overwijk:2022} offers an extensive collection of 22 billion curated web pages, previously utilized in TREC tracks such as the TREC Conversational Assistance Track (CAsT) \cite{owoicho:2022} and the forthcoming TREC Interactive Knowledge Assistance Track (iKAT) \cite{aliannejadi:2024}.
Another alternative is the MS MARCO V2 document collection, used in the TREC Deep Learning (DL) track.

\paragraph{Topic selection.} Recently, there has been a surge in datasets providing topics with long-form answers for evaluating RAG systems. ASQA \cite{stelmark:2022}, ELI5 \cite{fan:2019}, and QAMPARI~\cite{amouyal:2024} were utilized for evaluation in the Automatic LLMs’ Citation Evaluation (ALCE) framework \cite{gao:2023}. Similarly, related long-form QA datasets include AquaMuse~\cite{kulkarni:2020}, ExpertQA~\cite{malaviya:2023}, and TruthfulQA~\cite{lin:2022}.
Another recently introduced dataset is ClapNQ~\cite{rosenthal:2024}, created from the subset of Natural Questions (NQ)~\cite{kwiatkowski:2019} and HAGRID~\cite{kamalloo:2023} built on a subset of MS MARCO Dev \cite{bajaj:2016}.
Almost all previous datasets are built on English Wikipedia. In contrast, our work deliberately avoids English Wikipedia to prevent the overfitting commonly seen in existing benchmarks~\cite{thakur:2021, muennighoff:2023}.
In our work, we re-utilize topics from previous TREC tracks such as the Deep Learning (DL) track, because human judgments are available on the MS MARCO V2 corpora and Researchy Questions \cite{rosset:2024} as it covers a wide range of topics based on ClueWeb22~\cite{overwijk:2022}.

\section{Our Framework}
\ragnarok is an open-source, reproducible, and reusable framework implementing an end-to-end retrieval-augmented generation (RAG) pipeline, comprising two modules applied sequentially: (1) (R) retrieval and (2) (AG) augmented generation. 
Through the \ragnarok framework, we will provide several baselines to all participants in the upcoming TREC 2024 RAG track. 
An overview of the framework is provided in \autoref{fig:ragnarok-framework}. 
We first describe both modules and expand on the I/O specifications in our framework.

\paragraph{Retrieval Module} This module retrieves the relevant segments for a user topic as the input. It supports (i) first-stage lexical retrieval models such as BM25 \cite{robertson:2009} and (ii) reranking models such as RankZephyr~\cite{pradeep:2023b}. The retrieval system searches for relevant segments in the document collection and retrieves the top-$100$ segments further reranked by the reranker model to filter out the top-$20$ relevant segments for the next stage.

\paragraph{Augmented Generation Module} This module takes in the user topic and the top-$20$ retrieved segments (from the retrieval module) as the input and a prompting strategy to the large language model (LLM) to generate the answer response with in-context citations for the topic.
The answer response is divided into individual sentences, each sentence within the answer contains text and is grounded on retrieved documents provided as references.

\subsection{RAG Input/Output Definitions}

\paragraph{RAG Input} The input specifications are straightforward as the user can formulate any question they wish to ask, provide the user topic, and call our \ragnarok REST-API framework.

\paragraph{RAG Output} The user receives a JSON output in response to their topic from the \ragnarok framework. The first key in the output JSON schema, \texttt{references}, provides an ordered list of the top-$20$ ranked segment IDs from our retrieval module. Next,\texttt{answer}, provides the LLM-generated RAG answer to the user topic, presented as a top-to-bottom list of sentence-level texts with corresponding segment citations. All citations are zero-based indexed indicating the exact position of the segment ID from the \texttt{references} list. Finally, \texttt{response\_length}, provides the total count of the text characters present in the output RAG answer.

\begin{table}[t]
\setlength{\tabcolsep}{2pt}
\caption{Comparison of document and segment counts between versions V2 and V2.1 (our version after removing near-duplicates) of the MS MARCO collection.}
\vspace{-1em}
    \renewcommand{\arraystretch}{1}
    \small
    {\begin{tabular}{c@{\hspace{8\tabcolsep}} c@{\hspace{8\tabcolsep}} c}
    \toprule
    \textbf{Collection} & \textbf{Version V2} & \textbf{Version V2.1 (Ours)} \\ \midrule
    MS MARCO Document & 11,959,635 & 10,960,555 \\
    MS MARCO Segment &  124,131,414	& 113,520,750 \\
    \bottomrule
\end{tabular}}\label{tab:document-counts}
\end{table}

\section{Document Collection}\label{sec:document-collection}
The MS MARCO V2 document collection, earlier used in the TREC-DL tracks, contains a substantial overlap of near-duplicates (documents with sufficiently similar text information) within the collection \cite{craswell:2022,craswell:2024}. 
When left intact, these near-duplicates degrade the downstream retrieval accuracy and reduce the diversity of the collected documents, potentially impacting the effectiveness of RAG systems.
In addition, chunking, which breaks down a long verbose document into smaller compact representations is a key challenge in RAG, as the retrieved chunk representations correlate with the RAG answer quality~\cite{liu:2024}.

\paragraph{MS MARCO V2.1 Document Collection} We conduct a deduplication strategy in the MS MARCO V2 document collection to avoid near-duplicates in two stages. In the first stage, we establish an equivalence class of the documents using Locality Sensitive Hashing (LSH) with MinHash \cite{broder:1997} and 9-gram shingles.
Next, we selected a representative document for each equivalence class for our refined MS MARCO V2.1 document collection,\footnote{MS MARCO V2.1 document collection: \href{https://msmarco.z22.web.core.windows.net/msmarcoranking/msmarco\_v2.1\_doc.tar}{msmarco\_v2.1\_doc.tar}.} reducing the duplicates in the original MS MARCO V2 document collection by 8.35\% as shown in \autoref{tab:document-counts}.

\paragraph{MS MARCO V2.1 Segment Collection.} We segment the MS MARCO V2.1 document collection into overlapping segments (or chunks) and develop the MS MARCO V2.1 segment collection\footnote{MS MARCO V2.1 segment collection: \href{https://msmarco.z22.web.core.windows.net/msmarcoranking/msmarco\_v2.1\_doc\_segmented.tar}{msmarco\_v2.1\_doc\_segmented.tar}.} with more than 113 million text segments (\autoref{tab:document-counts}). We utilize a sliding window technique to generate the segments, by fixing the sliding window size of 10 sentences and a stride of 5 sentences to create each segment, roughly on average, between 500--1000 characters long. To easily map each segment back to the document, every segment contains the document ID within the segment ID. Further, two new fields: \texttt{start\_char} and \texttt{end\_char} indicate the start and the end position character of where the segment begins and ends in the mapped MS MARCO V2.1 document collection.

\section{Topic Collection}
Topics, i.e., user queries, are crucial for robust evaluation of RAG systems. Traditionally, popular retrieval and traditional QA benchmarks primarily consist of factoid queries, where answers are typically found within a single sentence or paragraph. However, these topics lack complexity, leading to short answers that can be easily memorized by LLMs. For instance, MS MARCO \cite{bajaj:2016} surprisingly contains up to 55\% factoid queries \cite{bolotova:2022, rosset:2024}.
To avoid short-form answers in RAG, we utilize two collections containing non-factoid topics covering information about diverse topics and requiring long-form answers. We describe these collections below:

\begin{table}[t]
\setlength{\tabcolsep}{1pt}
\caption{TREC-RAGgy and TREC-Researchy 2024 topic distribution. The table shows the top-$5$ categories in topic classification for TREC-RAGgy, intrinsic attributes for TREC-Researchy, and the first word in all topics. Definitions in more detail can be found in \autoref{sec:trec-raggy-appendix} \& {\ref{sec:trec-researchy-appendix}}.}
\vspace{-1em}
    \resizebox{0.48\textwidth}{!}{\renewcommand{\arraystretch}{1}
    \small
    {\begin{tabular}{l c@{\hspace{5\tabcolsep}} l c@{\hspace{10\tabcolsep}} l c@{\hspace{5\tabcolsep}} l c@{\hspace{5\tabcolsep}} l}
    \toprule
    \multicolumn{2}{l}{\textbf{TREC-RAGgy 2024}} & \multicolumn{2}{l}{} & \multicolumn{2}{l}{\textbf{TREC-Researchy 2024}} \\ \cmidrule(lr){1-4} \cmidrule(lr){5-8}
    \textbf{Topic Category} & \% & \textbf{First Word} & \% & \textbf{Intrinsic Attributes} & \% & \textbf{First Word} & \% \\ \midrule
    Aggregation & 24.2 & What & 37.5 & Knowledge-Intensive & 79.8 & How & 41.0 \\ 
    Simple w/ cond. & 23.3 & How & 27.5 & Multi-Faceted & 75.7 & Why & 25.5 \\
    Set & 20.8 & Why & 3.3 & Reasoning-Intensive & 75.5 & What & 15.0 \\ 
    Simple & 10.0 & Is & 2.5 & Subjective & 48.5 & Is & 5.2 \\ 
    Comparison & 6.7 & When & 1.7 & Assumptive & 25.7 & Should & 2.2 \\ \bottomrule
\end{tabular}}}\label{tab:merged-query-distribution}
\end{table}

\paragraph{TREC-RAGgy 2024}\label{sec:trec-raggy-2024}
We develop TREC-RAGgy 2024, a collection with topics filtered from TREC Deep Learning 2021-2023 tracks \cite{craswell:2021,craswell:2022,craswell:2024}, based on topic category and generated-answer classification. We classify each available topic into seven categories described in \autoref{sec:trec-raggy-appendix} and filter out a subset of topics that either have a long-form answer or require information aggregation across multiple sources of information. 
Out of the 210 original topics available, we filter and include 120 topics (57.1\%) in the TREC-RAGgy 2024 topic collection.\footnote{TREC-RAGgy 2024 topic collection: \href{https://github.com/castorini/anserini-tools/blob/master/topics-and-qrels/topics.rag24.raggy-dev.txt}{topics.rag24.raggy-dev.txt}.} From ~\autoref{tab:merged-query-distribution}, we observe 24.2\% of the topics included are ``aggregation'', indicating RAG systems require to aggregate information from multiple retrieved segments to generate an accurate long-form answer. Similarly, 65\% of the topics start with ``what'' or ``how''. Overall, a majority of the topics are useful for evaluation containing diverse topic categories requiring a long-form answer.

\paragraph{TREC-Researchy 2024}\label{sec:researchy-questions}
Researchy Questions, introduced in \citet{rosset:2024}, contains 102K non-factoid topics with long-form answers. These topics were curated from Bing Search logs and evaluated by GPT-4 on a scale of 0--10 based on eight intrinsic attributes, such as subjectivity and multifacetedness (definitions provided in \autoref{sec:trec-researchy-appendix}). Notably, unlike TREC-RAGgy 2024, these queries lack relevance judgments.
To curate a smaller development subset for a faster evaluation of RAG systems, we employ a sampler designed to maximize diversity based on the eight intrinsic attributes. 
This is achieved by iteratively selecting the query with the highest $l_1$ norm in the intrinsic attribute space (of all eight dimensions) relative to the already-sampled set.
The resultant topic set we dub as TREC-Researchy 2024.\footnote{TREC-Researchy 2024 topic collection: \href{https://github.com/castorini/anserini-tools/blob/master/topics-and-qrels/topics.rag24.researchy-dev.txt}{topics.rag24.researchy-dev.txt}} 
% serves as an additional development set for the TREC 2024 RAG track.
From \autoref{tab:merged-query-distribution}, about $80$\% of the topics are Knowledge-Intensive and about $76$\% are Multi-Faceted highlighting the need for effective RAG systems. 
Additionally, 66.5\% of topics start with ``how'' or ``why'', emphasizing explanatory questions. 
These distributions suggest that TREC-Researchy 2024 prioritizes complex and multi-dimensional topics.

\section{TREC 2024 RAG Baselines}\label{sec:baselines}
\paragraph{Retrieval.} Our retrieval module integrates both first-stage retrievers and rerankers.
We use BM25 available in Anserini~\cite{anserini} with the following default parameters ($k_1 = 0.9$ and $b= 0.4$), to retrieve the top-$100$ segments for a given topic. 
Next, RankZephyr~\cite{pradeep:2023b}, a state-of-the-art listwise reranker, is used to rerank the top-$100$ candidates.
We use RankZephyr\textsubscript{$\rho$}, a variant, that reranks the candidates progressively, i.e., makes three passes iteratively, refining the final ranked candidate list to achieve better precision.
An easy-to-use implementation of RankZephyr is available via the \texttt{rank\_llm} package, along with various other rerankers like RankGPT~\cite{rankgpt}, which we provide as secondary baselines.
Finally, the top-$20$ reranked documents from the document collection are passed onto the next stage for RAG generation.

\paragraph{Augmented Generation.} Our generation module integrates two popular and commercially available LLMs: (i) Command R+ is Cohere's instruction following LLM developed for complex RAG pipelines~\cite{command_r_plus}; (ii) GPT-4o is the latest GPT version from OpenAI~\cite{openai_gpt4o}. Given that Command R+ cites in a span level, we map the citations to their parent sentences.
For GPT-4o, we follow the zero-shot ChatQA prompt template~\cite{liu:2024b} and cite relevant segments within the text (in-line) using the IEEE format. An example of the prompt template is shown in \autoref{fig:prompt}, in the Appendix.

\paragraph{RAG-Bench Evaluation}
Evaluating different RAG answers is challenging as multiple factors within the output response are crucial for effectiveness evaluation. To combat this, recent works rely on an LLM-as-a-judge setup \cite{zheng:2024}, where strong LLM assessors judge the RAG-generated output in a pairwise evaluation style (side-by-side) in a head-on tournament. In our work, we briefly overview our baseline techniques using human evaluators. A complete illustration can be found in \autoref{tab:e2e-trec-raggy}, in the Appendix. The Command R+ baseline outputs shorter answers and cites more relevant segments, whereas, the GPT-4o baseline outputs longer and more detailed answers and cites fewer segments. Therefore, for topics in both TREC-Raggy and TREC-Researchy 2024, GPT-4o intuitively is the better choice for RAG answer generation. We leave it for future work, to empirically compute the win rates (in \%) between our baselines in the RAG-bench evaluation.

\subsection{\ragnarok System Arena}\label{sec:rag_system_Arena}
Heavily inspired by the success of Chatbot Arena~\cite{chiang:2024, zheng:2024}, a crowdsourcing benchmark WebUI featuring anonymous battles, we extend the concept to multi-stage configurable RAG pipelines with \ragnarok.
In the arena, users interact with two unblinded/blinded RAG systems simultaneously, issuing the same topic to both.
The participants evaluate and select the pipeline that delivers their most preferred response, with the identities of the modules in the end-to-end pipeline revealed after the voting process in the blinded case.
We leverage Gradio~\cite{abid:2019} to build the WebUI for \ragnarok.
Each step of the pipeline uses REST APIs for intercommunication, enabling easy module switching within the pipeline.
This modular design simplifies the integration of different retrieval and LLM configurations, enhancing scalability and maintainability.

\autoref{fig:ragnarok-battle} in the Appendix illustrates an example topic ``what inspired pink floyd’s the wall?'' processed by two different pipelines: Pipeline A, comprising BM25 $\rightarrow$ RankZephyr $\rightarrow$ GPT-4o (left), and Pipeline B, comprising BM25 $\rightarrow$ RankGPT-4o $\rightarrow$ Command R+ (right) in the unblinded tab. 
The outputs generated by each pipeline are compared, allowing users to discern which system provided a more satisfactory answer.
Note that when the user hovers the mouse over a citation, they can preview the cited segment.
Further, in Appendix \ref{app:system_arena}, we discuss the blinded pairwise evaluation and the responses (JSON output) tab, available in the WebUI for \ragnarok.

\section{Ongoing Work}
\ragnarok is the first step for the ongoing work in the TREC 2024 RAG track, 
by releasing the document collections, development topics, and baseline strategies for participants. 
We will continue to update the pipelines to include more diverse retrieval models including state-of-the-art dual encoders such as Artic-Embed \cite{merrick:2024} and effective pointwise/pairwise rerankers~\cite{EMD}.
We plan to add additional support for more advanced RAG techniques like SelfRAG~\cite{asai:2023} and CRAG~\cite{yan:2024}.
For the TREC 2024 RAG track test topics, we plan to conduct a new and fresh scrape of the Bing search logs closer to the submission period. This approach will compile a fresh and recent set of topics, similar to \citet{rosset:2024}, thereby minimizing the risk of data leakage and ensuring a fair evaluation with existing commercially available LLMs.

The next phase of our efforts will focus on finalizing the evaluation methodology using an automatic nugget-based evaluation, following earlier work in \citet{lin:2006} and first discussed in the TREC RAG 2024 presentation deck.\footnote{\href{https://cs.uwaterloo.ca/\~jimmylin/publications/Lin\_etal\_TREC2023-planning.pdf}{https://cs.uwaterloo.ca/\~jimmylin/publications/Lin\_etal\_TREC2023-planning.pdf}} 
% An information nugget \cite{} is a short answer or fact present in the collection and addresses an aspect of the RAG answer. 
The nugget-based evaluation is recently gaining popularity \cite{alaofi:2024, raina:2024, arabzadeh:2024, mayfield:2024}, and is becoming the de facto strategy for RAG evaluation.

\section{Conclusion}
The emergence of retrieval-augmented generation (RAG) has revolutionized modern search systems by allowing real-time data incorporation into large language models (LLMs). 
In our work, we develop a reusable end-to-end framework, \ragnarok, providing reproducible baselines and a WebUI serving a RAG battle arena for retriever, reranker, and generation models.
We also introduce the MS MARCO V2.1 collection, carefully curated topics from the TREC-DL 2021-2023 queries and Researchy Questions, and I/O definitions to assist users in the RAG paradigm.
Additionally, the paper identifies key industrial baselines (such as Cohere’s Command R+ and OpenAI’s GPT-4o) and includes a qualitative analysis of the baselines on the development topics. 
By open-sourcing this framework, we aim to standardize RAG applications in preparation for the upcoming TREC 2024 RAG challenge.

%%
%% The acknowledgments section is defined using the "acks" environment
%% (and NOT an unnumbered section). This ensures the proper
%% identification of the section in the article metadata, and the
%% consistent spelling of the heading.
\begin{acks}
We thank Ian Soboroff for the MS MARCO V2 document collection deduplication for our TREC 2024 RAG track, Cohere for providing us with the necessary credits to evaluate Command-R+, and Microsoft for providing Azure credits to evaluate GPT-4o. Additionally, we thank Corby Rosset for the discussions surrounding Researchy Questions~\cite{rosset:2024}. 
\end{acks}

%%
%% The next two lines define the bibliography style to be used, and
%% the bibliography file.
\bibliographystyle{ACM-Reference-Format}
\bibliography{ref}

%%% -*-BibTeX-*-
%%% Do NOT edit. File created by BibTeX with style
%%% ACM-Reference-Format-Journals [18-Jan-2012].

\begin{thebibliography}{58}

%%% ====================================================================
%%% NOTE TO THE USER: you can override these defaults by providing
%%% customized versions of any of these macros before the \bibliography
%%% command.  Each of them MUST provide its own final punctuation,
%%% except for \shownote{}, \showDOI{}, and \showURL{}.  The latter two
%%% do not use final punctuation, in order to avoid confusing it with
%%% the Web address.
%%%
%%% To suppress output of a particular field, define its macro to expand
%%% to an empty string, or better, \unskip, like this:
%%%
%%% \newcommand{\showDOI}[1]{\unskip}   % LaTeX syntax
%%%
%%% \def \showDOI #1{\unskip}           % plain TeX syntax
%%%
%%% ====================================================================

\ifx \showCODEN    \undefined \def \showCODEN     #1{\unskip}     \fi
\ifx \showDOI      \undefined \def \showDOI       #1{#1}\fi
\ifx \showISBNx    \undefined \def \showISBNx     #1{\unskip}     \fi
\ifx \showISBNxiii \undefined \def \showISBNxiii  #1{\unskip}     \fi
\ifx \showISSN     \undefined \def \showISSN      #1{\unskip}     \fi
\ifx \showLCCN     \undefined \def \showLCCN      #1{\unskip}     \fi
\ifx \shownote     \undefined \def \shownote      #1{#1}          \fi
\ifx \showarticletitle \undefined \def \showarticletitle #1{#1}   \fi
\ifx \showURL      \undefined \def \showURL       {\relax}        \fi
% The following commands are used for tagged output and should be
% invisible to TeX
\providecommand\bibfield[2]{#2}
\providecommand\bibinfo[2]{#2}
\providecommand\natexlab[1]{#1}
\providecommand\showeprint[2][]{arXiv:#2}

\bibitem[Abid et~al\mbox{.}(2019)]%
        {abid:2019}
\bibfield{author}{\bibinfo{person}{Abubakar Abid}, \bibinfo{person}{Ali Abdalla}, \bibinfo{person}{Ali Abid}, \bibinfo{person}{Dawood Khan}, \bibinfo{person}{Abdulrahman Alfozan}, {and} \bibinfo{person}{James~Y. Zou}.} \bibinfo{year}{2019}\natexlab{}.
\newblock \showarticletitle{Gradio: Hassle-Free Sharing and Testing of {ML} Models in the Wild}.
\newblock \bibinfo{journal}{\emph{CoRR}}  \bibinfo{volume}{abs/1906.02569} (\bibinfo{year}{2019}).
\newblock
\showeprint[arXiv]{1906.02569}
\urldef\tempurl%
\url{http://arxiv.org/abs/1906.02569}
\showURL{%
\tempurl}


\bibitem[Alaofi et~al\mbox{.}(2024)]%
        {alaofi:2024}
\bibfield{author}{\bibinfo{person}{Marwah Alaofi}, \bibinfo{person}{Negar Arabzadeh}, \bibinfo{person}{Charles L.~A. Clarke}, {and} \bibinfo{person}{Mark Sanderson}.} \bibinfo{year}{2024}\natexlab{}.
\newblock \showarticletitle{Generative Information Retrieval Evaluation}.
\newblock \bibinfo{journal}{\emph{CoRR}}  \bibinfo{volume}{abs/2404.08137} (\bibinfo{year}{2024}).
\newblock
\urldef\tempurl%
\url{https://doi.org/10.48550/ARXIV.2404.08137}
\showDOI{\tempurl}
\showeprint[arXiv]{2404.08137}


\bibitem[Aliannejadi et~al\mbox{.}(2024)]%
        {aliannejadi:2024}
\bibfield{author}{\bibinfo{person}{Mohammad Aliannejadi}, \bibinfo{person}{Zahra Abbasiantaeb}, \bibinfo{person}{Shubham Chatterjee}, \bibinfo{person}{Jeffery Dalton}, {and} \bibinfo{person}{Leif Azzopardi}.} \bibinfo{year}{2024}\natexlab{}.
\newblock \showarticletitle{{TREC} iKAT 2023: The Interactive Knowledge Assistance Track Overview}.
\newblock \bibinfo{journal}{\emph{CoRR}}  \bibinfo{volume}{abs/2401.01330} (\bibinfo{year}{2024}).
\newblock
\urldef\tempurl%
\url{https://doi.org/10.48550/ARXIV.2401.01330}
\showDOI{\tempurl}
\showeprint[arXiv]{2401.01330}


\bibitem[Amouyal et~al\mbox{.}(2022)]%
        {amouyal:2024}
\bibfield{author}{\bibinfo{person}{Samuel~Joseph Amouyal}, \bibinfo{person}{Ohad Rubin}, \bibinfo{person}{Ori Yoran}, \bibinfo{person}{Tomer Wolfson}, \bibinfo{person}{Jonathan Herzig}, {and} \bibinfo{person}{Jonathan Berant}.} \bibinfo{year}{2022}\natexlab{}.
\newblock \showarticletitle{{QAMPARI:} : An Open-domain Question Answering Benchmark for Questions with Many Answers from Multiple Paragraphs}.
\newblock \bibinfo{journal}{\emph{CoRR}}  \bibinfo{volume}{abs/2205.12665} (\bibinfo{year}{2022}).
\newblock
\urldef\tempurl%
\url{https://doi.org/10.48550/ARXIV.2205.12665}
\showDOI{\tempurl}
\showeprint[arXiv]{2205.12665}


\bibitem[Anil et~al\mbox{.}(2023)]%
        {anil:2023}
\bibfield{author}{\bibinfo{person}{Rohan Anil}, \bibinfo{person}{Sebastian Borgeaud}, \bibinfo{person}{Yonghui Wu}, \bibinfo{person}{Jean{-}Baptiste Alayrac}, \bibinfo{person}{Jiahui Yu}, \bibinfo{person}{Radu Soricut}, \bibinfo{person}{Johan Schalkwyk}, \bibinfo{person}{Andrew~M. Dai}, \bibinfo{person}{Anja Hauth}, \bibinfo{person}{Katie Millican}, \bibinfo{person}{David Silver}, \bibinfo{person}{Slav Petrov}, \bibinfo{person}{Melvin Johnson}, \bibinfo{person}{Ioannis Antonoglou}, \bibinfo{person}{Julian Schrittwieser}, \bibinfo{person}{Amelia Glaese}, \bibinfo{person}{Jilin Chen}, \bibinfo{person}{Emily Pitler}, \bibinfo{person}{Timothy~P. Lillicrap}, \bibinfo{person}{Angeliki Lazaridou}, \bibinfo{person}{Orhan Firat}, \bibinfo{person}{James Molloy}, \bibinfo{person}{Michael Isard}, \bibinfo{person}{Paul~Ronald Barham}, \bibinfo{person}{Tom Hennigan}, \bibinfo{person}{Benjamin Lee}, \bibinfo{person}{Fabio Viola}, \bibinfo{person}{Malcolm Reynolds}, \bibinfo{person}{Yuanzhong Xu}, \bibinfo{person}{Ryan
  Doherty}, \bibinfo{person}{Eli Collins}, \bibinfo{person}{Clemens Meyer}, \bibinfo{person}{Eliza Rutherford}, \bibinfo{person}{Erica Moreira}, \bibinfo{person}{Kareem Ayoub}, \bibinfo{person}{Megha Goel}, \bibinfo{person}{George Tucker}, \bibinfo{person}{Enrique Piqueras}, \bibinfo{person}{Maxim Krikun}, \bibinfo{person}{Iain Barr}, \bibinfo{person}{Nikolay Savinov}, \bibinfo{person}{Ivo Danihelka}, \bibinfo{person}{Becca Roelofs}, \bibinfo{person}{Ana{\"{\i}}s White}, \bibinfo{person}{Anders Andreassen}, \bibinfo{person}{Tamara von Glehn}, \bibinfo{person}{Lakshman Yagati}, \bibinfo{person}{Mehran Kazemi}, \bibinfo{person}{Lucas Gonzalez}, \bibinfo{person}{Misha Khalman}, \bibinfo{person}{Jakub Sygnowski}, {and} \bibinfo{person}{et al.}} \bibinfo{year}{2023}\natexlab{}.
\newblock \showarticletitle{Gemini: {A} Family of Highly Capable Multimodal Models}.
\newblock \bibinfo{journal}{\emph{CoRR}}  \bibinfo{volume}{abs/2312.11805} (\bibinfo{year}{2023}).
\newblock
\urldef\tempurl%
\url{https://doi.org/10.48550/ARXIV.2312.11805}
\showDOI{\tempurl}
\showeprint[arXiv]{2312.11805}


\bibitem[Arabzadeh and Clarke(2024)]%
        {arabzadeh:2024}
\bibfield{author}{\bibinfo{person}{Negar Arabzadeh} {and} \bibinfo{person}{Charles L.~A. Clarke}.} \bibinfo{year}{2024}\natexlab{}.
\newblock \showarticletitle{A Comparison of Methods for Evaluating Generative {IR}}.
\newblock \bibinfo{journal}{\emph{CoRR}}  \bibinfo{volume}{abs/2404.04044} (\bibinfo{year}{2024}).
\newblock
\urldef\tempurl%
\url{https://doi.org/10.48550/ARXIV.2404.04044}
\showDOI{\tempurl}
\showeprint[arXiv]{2404.04044}


\bibitem[Asai et~al\mbox{.}(2023)]%
        {asai:2023}
\bibfield{author}{\bibinfo{person}{Akari Asai}, \bibinfo{person}{Zeqiu Wu}, \bibinfo{person}{Yizhong Wang}, \bibinfo{person}{Avirup Sil}, {and} \bibinfo{person}{Hannaneh Hajishirzi}.} \bibinfo{year}{2023}\natexlab{}.
\newblock \showarticletitle{Self-RAG: Learning to Retrieve, Generate, and Critique through Self-Reflection}.
\newblock \bibinfo{journal}{\emph{CoRR}}  \bibinfo{volume}{abs/2310.11511} (\bibinfo{year}{2023}).
\newblock
\urldef\tempurl%
\url{https://doi.org/10.48550/ARXIV.2310.11511}
\showDOI{\tempurl}
\showeprint[arXiv]{2310.11511}


\bibitem[Bajaj et~al\mbox{.}(2016)]%
        {bajaj:2016}
\bibfield{author}{\bibinfo{person}{Payal Bajaj}, \bibinfo{person}{Daniel Campos}, \bibinfo{person}{Nick Craswell}, \bibinfo{person}{Li Deng}, \bibinfo{person}{Jianfeng Gao}, \bibinfo{person}{Xiaodong Liu}, \bibinfo{person}{Rangan Majumder}, \bibinfo{person}{Andrew McNamara}, \bibinfo{person}{Bhaskar Mitra}, \bibinfo{person}{Tri Nguyen}, \bibinfo{person}{Mir Rosenberg}, \bibinfo{person}{Xia Song}, \bibinfo{person}{Alina Stoica}, \bibinfo{person}{Saurabh Tiwary}, {and} \bibinfo{person}{Tong Wang}.} \bibinfo{year}{2016}\natexlab{}.
\newblock \showarticletitle{{MS} {MARCO:} {A} Human Generated MAchine Reading COmprehension Dataset}.
\newblock \bibinfo{journal}{\emph{CoRR}}  \bibinfo{volume}{abs/1611.09268} (\bibinfo{year}{2016}).
\newblock
\showeprint[arXiv]{1611.09268}
\urldef\tempurl%
\url{http://arxiv.org/abs/1611.09268}
\showURL{%
\tempurl}


\bibitem[Bolotova et~al\mbox{.}(2022)]%
        {bolotova:2022}
\bibfield{author}{\bibinfo{person}{Valeria Bolotova}, \bibinfo{person}{Vladislav Blinov}, \bibinfo{person}{Falk Scholer}, \bibinfo{person}{W.~Bruce Croft}, {and} \bibinfo{person}{Mark Sanderson}.} \bibinfo{year}{2022}\natexlab{}.
\newblock \showarticletitle{A Non-Factoid Question-Answering Taxonomy}. In \bibinfo{booktitle}{\emph{{SIGIR} '22: The 45th International {ACM} {SIGIR} Conference on Research and Development in Information Retrieval, Madrid, Spain, July 11 - 15, 2022}}, \bibfield{editor}{\bibinfo{person}{Enrique Amig{\'{o}}}, \bibinfo{person}{Pablo Castells}, \bibinfo{person}{Julio Gonzalo}, \bibinfo{person}{Ben Carterette}, \bibinfo{person}{J.~Shane Culpepper}, {and} \bibinfo{person}{Gabriella Kazai}} (Eds.). \bibinfo{publisher}{{ACM}}, \bibinfo{pages}{1196--1207}.
\newblock
\urldef\tempurl%
\url{https://doi.org/10.1145/3477495.3531926}
\showDOI{\tempurl}


\bibitem[Borgeaud et~al\mbox{.}(2022)]%
        {borgeaud:2022}
\bibfield{author}{\bibinfo{person}{Sebastian Borgeaud}, \bibinfo{person}{Arthur Mensch}, \bibinfo{person}{Jordan Hoffmann}, \bibinfo{person}{Trevor Cai}, \bibinfo{person}{Eliza Rutherford}, \bibinfo{person}{Katie Millican}, \bibinfo{person}{George van~den Driessche}, \bibinfo{person}{Jean{-}Baptiste Lespiau}, \bibinfo{person}{Bogdan Damoc}, \bibinfo{person}{Aidan Clark}, \bibinfo{person}{Diego de Las~Casas}, \bibinfo{person}{Aurelia Guy}, \bibinfo{person}{Jacob Menick}, \bibinfo{person}{Roman Ring}, \bibinfo{person}{Tom Hennigan}, \bibinfo{person}{Saffron Huang}, \bibinfo{person}{Loren Maggiore}, \bibinfo{person}{Chris Jones}, \bibinfo{person}{Albin Cassirer}, \bibinfo{person}{Andy Brock}, \bibinfo{person}{Michela Paganini}, \bibinfo{person}{Geoffrey Irving}, \bibinfo{person}{Oriol Vinyals}, \bibinfo{person}{Simon Osindero}, \bibinfo{person}{Karen Simonyan}, \bibinfo{person}{Jack~W. Rae}, \bibinfo{person}{Erich Elsen}, {and} \bibinfo{person}{Laurent Sifre}.} \bibinfo{year}{2022}\natexlab{}.
\newblock \showarticletitle{Improving Language Models by Retrieving from Trillions of Tokens}. In \bibinfo{booktitle}{\emph{International Conference on Machine Learning, {ICML} 2022, 17-23 July 2022, Baltimore, Maryland, {USA}}} \emph{(\bibinfo{series}{Proceedings of Machine Learning Research}, Vol.~\bibinfo{volume}{162})}, \bibfield{editor}{\bibinfo{person}{Kamalika Chaudhuri}, \bibinfo{person}{Stefanie Jegelka}, \bibinfo{person}{Le~Song}, \bibinfo{person}{Csaba Szepesv{\'{a}}ri}, \bibinfo{person}{Gang Niu}, {and} \bibinfo{person}{Sivan Sabato}} (Eds.). \bibinfo{publisher}{{PMLR}}, \bibinfo{pages}{2206--2240}.
\newblock
\urldef\tempurl%
\url{https://proceedings.mlr.press/v162/borgeaud22a.html}
\showURL{%
\tempurl}


\bibitem[Broder(1997)]%
        {broder:1997}
\bibfield{author}{\bibinfo{person}{Andrei~Z. Broder}.} \bibinfo{year}{1997}\natexlab{}.
\newblock \showarticletitle{On the resemblance and containment of documents}. In \bibinfo{booktitle}{\emph{Compression and Complexity of {SEQUENCES} 1997, Positano, Amalfitan Coast, Salerno, Italy, June 11-13, 1997, Proceedings}}, \bibfield{editor}{\bibinfo{person}{Bruno Carpentieri}, \bibinfo{person}{Alfredo~De Santis}, \bibinfo{person}{Ugo Vaccaro}, {and} \bibinfo{person}{James~A. Storer}} (Eds.). \bibinfo{publisher}{{IEEE}}, \bibinfo{pages}{21--29}.
\newblock
\urldef\tempurl%
\url{https://doi.org/10.1109/SEQUEN.1997.666900}
\showDOI{\tempurl}


\bibitem[Chase(2022)]%
        {langchain}
\bibfield{author}{\bibinfo{person}{Harrison Chase}.} \bibinfo{year}{2022}\natexlab{}.
\newblock \bibinfo{booktitle}{\emph{LangChain}}.
\newblock
\urldef\tempurl%
\url{https://github.com/langchain-ai/langchain}
\showURL{%
\tempurl}


\bibitem[Chiang et~al\mbox{.}(2024)]%
        {chiang:2024}
\bibfield{author}{\bibinfo{person}{Wei{-}Lin Chiang}, \bibinfo{person}{Lianmin Zheng}, \bibinfo{person}{Ying Sheng}, \bibinfo{person}{Anastasios~Nikolas Angelopoulos}, \bibinfo{person}{Tianle Li}, \bibinfo{person}{Dacheng Li}, \bibinfo{person}{Hao Zhang}, \bibinfo{person}{Banghua Zhu}, \bibinfo{person}{Michael~I. Jordan}, \bibinfo{person}{Joseph~E. Gonzalez}, {and} \bibinfo{person}{Ion Stoica}.} \bibinfo{year}{2024}\natexlab{}.
\newblock \showarticletitle{Chatbot Arena: An Open Platform for Evaluating LLMs by Human Preference}.
\newblock \bibinfo{journal}{\emph{CoRR}}  \bibinfo{volume}{abs/2403.04132} (\bibinfo{year}{2024}).
\newblock
\urldef\tempurl%
\url{https://doi.org/10.48550/ARXIV.2403.04132}
\showDOI{\tempurl}
\showeprint[arXiv]{2403.04132}


\bibitem[Cohere(2024)]%
        {command_r_plus}
\bibfield{author}{\bibinfo{person}{Cohere}.} \bibinfo{year}{2024}\natexlab{}.
\newblock \bibinfo{booktitle}{\emph{Introducing Command R+: A Scalable LLM Built for Business}}.
\newblock
\urldef\tempurl%
\url{https://cohere.com/blog/command-r-plus-microsoft-azure}
\showURL{%
\tempurl}


\bibitem[Craswell et~al\mbox{.}(2021)]%
        {craswell:2021}
\bibfield{author}{\bibinfo{person}{Nick Craswell}, \bibinfo{person}{Bhaskar Mitra}, \bibinfo{person}{Emine Yilmaz}, \bibinfo{person}{Daniel Campos}, {and} \bibinfo{person}{Jimmy Lin}.} \bibinfo{year}{2021}\natexlab{}.
\newblock \showarticletitle{Overview of the {TREC} 2021 Deep Learning Track}. In \bibinfo{booktitle}{\emph{Proceedings of the Thirtieth Text REtrieval Conference, {TREC} 2021, online, November 15-19, 2021}} \emph{(\bibinfo{series}{{NIST} Special Publication}, Vol.~\bibinfo{volume}{500-335})}, \bibfield{editor}{\bibinfo{person}{Ian Soboroff} {and} \bibinfo{person}{Angela Ellis}} (Eds.). \bibinfo{publisher}{National Institute of Standards and Technology {(NIST)}}.
\newblock
\urldef\tempurl%
\url{https://trec.nist.gov/pubs/trec30/papers/Overview-DL.pdf}
\showURL{%
\tempurl}


\bibitem[Craswell et~al\mbox{.}(2022)]%
        {craswell:2022}
\bibfield{author}{\bibinfo{person}{Nick Craswell}, \bibinfo{person}{Bhaskar Mitra}, \bibinfo{person}{Emine Yilmaz}, \bibinfo{person}{Daniel Campos}, \bibinfo{person}{Jimmy Lin}, \bibinfo{person}{Ellen~M. Voorhees}, {and} \bibinfo{person}{Ian Soboroff}.} \bibinfo{year}{2022}\natexlab{}.
\newblock \showarticletitle{Overview of the {TREC} 2022 Deep Learning Track}. In \bibinfo{booktitle}{\emph{Proceedings of the Thirty-First Text REtrieval Conference, {TREC} 2022, online, November 15-19, 2022}} \emph{(\bibinfo{series}{{NIST} Special Publication}, Vol.~\bibinfo{volume}{500-338})}, \bibfield{editor}{\bibinfo{person}{Ian Soboroff} {and} \bibinfo{person}{Angela Ellis}} (Eds.). \bibinfo{publisher}{National Institute of Standards and Technology {(NIST)}}.
\newblock
\urldef\tempurl%
\url{https://trec.nist.gov/pubs/trec31/papers/Overview\_deep.pdf}
\showURL{%
\tempurl}


\bibitem[Craswell et~al\mbox{.}(2024)]%
        {craswell:2024}
\bibfield{author}{\bibinfo{person}{Nick Craswell}, \bibinfo{person}{Bhaskar Mitra}, \bibinfo{person}{Emine Yilmaz}, \bibinfo{person}{Hossein~A. Rahmani}, \bibinfo{person}{Daniel Campos}, \bibinfo{person}{Jimmy Lin}, \bibinfo{person}{Ellen~M. Voorhees}, {and} \bibinfo{person}{Ian Soboroff}.} \bibinfo{year}{2024}\natexlab{}.
\newblock \showarticletitle{Overview of the TREC 2023 Deep Learning Track}. In \bibinfo{booktitle}{\emph{Text REtrieval Conference (TREC)}}. NIST, \bibinfo{publisher}{TREC}.
\newblock
\urldef\tempurl%
\url{https://www.microsoft.com/en-us/research/publication/overview-of-the-trec-2023-deep-learning-track/}
\showURL{%
\tempurl}


\bibitem[Fan et~al\mbox{.}(2019)]%
        {fan:2019}
\bibfield{author}{\bibinfo{person}{Angela Fan}, \bibinfo{person}{Yacine Jernite}, \bibinfo{person}{Ethan Perez}, \bibinfo{person}{David Grangier}, \bibinfo{person}{Jason Weston}, {and} \bibinfo{person}{Michael Auli}.} \bibinfo{year}{2019}\natexlab{}.
\newblock \showarticletitle{{ELI5:} Long Form Question Answering}. In \bibinfo{booktitle}{\emph{Proceedings of the 57th Conference of the Association for Computational Linguistics, {ACL} 2019, Florence, Italy, July 28- August 2, 2019, Volume 1: Long Papers}}, \bibfield{editor}{\bibinfo{person}{Anna Korhonen}, \bibinfo{person}{David~R. Traum}, {and} \bibinfo{person}{Llu{\'{\i}}s M{\`{a}}rquez}} (Eds.). \bibinfo{publisher}{Association for Computational Linguistics}, \bibinfo{pages}{3558--3567}.
\newblock
\urldef\tempurl%
\url{https://doi.org/10.18653/V1/P19-1346}
\showDOI{\tempurl}


\bibitem[Gao et~al\mbox{.}(2023b)]%
        {gao:2023}
\bibfield{author}{\bibinfo{person}{Tianyu Gao}, \bibinfo{person}{Howard Yen}, \bibinfo{person}{Jiatong Yu}, {and} \bibinfo{person}{Danqi Chen}.} \bibinfo{year}{2023}\natexlab{b}.
\newblock \showarticletitle{Enabling Large Language Models to Generate Text with Citations}. In \bibinfo{booktitle}{\emph{Proceedings of the 2023 Conference on Empirical Methods in Natural Language Processing, {EMNLP} 2023, Singapore, December 6-10, 2023}}, \bibfield{editor}{\bibinfo{person}{Houda Bouamor}, \bibinfo{person}{Juan Pino}, {and} \bibinfo{person}{Kalika Bali}} (Eds.). \bibinfo{publisher}{Association for Computational Linguistics}, \bibinfo{pages}{6465--6488}.
\newblock
\urldef\tempurl%
\url{https://doi.org/10.18653/V1/2023.EMNLP-MAIN.398}
\showDOI{\tempurl}


\bibitem[Gao et~al\mbox{.}(2023a)]%
        {gao:2023b}
\bibfield{author}{\bibinfo{person}{Yunfan Gao}, \bibinfo{person}{Yun Xiong}, \bibinfo{person}{Xinyu Gao}, \bibinfo{person}{Kangxiang Jia}, \bibinfo{person}{Jinliu Pan}, \bibinfo{person}{Yuxi Bi}, \bibinfo{person}{Yi Dai}, \bibinfo{person}{Jiawei Sun}, \bibinfo{person}{Qianyu Guo}, \bibinfo{person}{Meng Wang}, {and} \bibinfo{person}{Haofen Wang}.} \bibinfo{year}{2023}\natexlab{a}.
\newblock \showarticletitle{Retrieval-Augmented Generation for Large Language Models: {A} Survey}.
\newblock \bibinfo{journal}{\emph{CoRR}}  \bibinfo{volume}{abs/2312.10997} (\bibinfo{year}{2023}).
\newblock
\urldef\tempurl%
\url{https://doi.org/10.48550/ARXIV.2312.10997}
\showDOI{\tempurl}
\showeprint[arXiv]{2312.10997}


\bibitem[Guu et~al\mbox{.}(2020)]%
        {guu:2020}
\bibfield{author}{\bibinfo{person}{Kelvin Guu}, \bibinfo{person}{Kenton Lee}, \bibinfo{person}{Zora Tung}, \bibinfo{person}{Panupong Pasupat}, {and} \bibinfo{person}{Ming{-}Wei Chang}.} \bibinfo{year}{2020}\natexlab{}.
\newblock \showarticletitle{Retrieval Augmented Language Model Pre-Training}. In \bibinfo{booktitle}{\emph{Proceedings of the 37th International Conference on Machine Learning, {ICML} 2020, 13-18 July 2020, Virtual Event}} \emph{(\bibinfo{series}{Proceedings of Machine Learning Research}, Vol.~\bibinfo{volume}{119})}. \bibinfo{publisher}{{PMLR}}, \bibinfo{pages}{3929--3938}.
\newblock
\urldef\tempurl%
\url{http://proceedings.mlr.press/v119/guu20a.html}
\showURL{%
\tempurl}


\bibitem[Izacard and Grave(2021)]%
        {izacard:2021}
\bibfield{author}{\bibinfo{person}{Gautier Izacard} {and} \bibinfo{person}{Edouard Grave}.} \bibinfo{year}{2021}\natexlab{}.
\newblock \showarticletitle{Leveraging Passage Retrieval with Generative Models for Open Domain Question Answering}. In \bibinfo{booktitle}{\emph{Proceedings of the 16th Conference of the European Chapter of the Association for Computational Linguistics: Main Volume, {EACL} 2021, Online, April 19 - 23, 2021}}, \bibfield{editor}{\bibinfo{person}{Paola Merlo}, \bibinfo{person}{J{\"{o}}rg Tiedemann}, {and} \bibinfo{person}{Reut Tsarfaty}} (Eds.). \bibinfo{publisher}{Association for Computational Linguistics}, \bibinfo{pages}{874--880}.
\newblock
\urldef\tempurl%
\url{https://doi.org/10.18653/V1/2021.EACL-MAIN.74}
\showDOI{\tempurl}


\bibitem[Jimeno{-}Yepes et~al\mbox{.}(2024)]%
        {yepes:2024}
\bibfield{author}{\bibinfo{person}{Antonio Jimeno{-}Yepes}, \bibinfo{person}{Yao You}, \bibinfo{person}{Jan Milczek}, \bibinfo{person}{Sebastian Laverde}, {and} \bibinfo{person}{Renyu Li}.} \bibinfo{year}{2024}\natexlab{}.
\newblock \showarticletitle{Financial Report Chunking for Effective Retrieval Augmented Generation}.
\newblock \bibinfo{journal}{\emph{CoRR}}  \bibinfo{volume}{abs/2402.05131} (\bibinfo{year}{2024}).
\newblock
\urldef\tempurl%
\url{https://doi.org/10.48550/ARXIV.2402.05131}
\showDOI{\tempurl}
\showeprint[arXiv]{2402.05131}


\bibitem[Jin et~al\mbox{.}(2024)]%
        {jin:2024}
\bibfield{author}{\bibinfo{person}{Jiajie Jin}, \bibinfo{person}{Yutao Zhu}, \bibinfo{person}{Xinyu Yang}, \bibinfo{person}{Chenghao Zhang}, {and} \bibinfo{person}{Zhicheng Dou}.} \bibinfo{year}{2024}\natexlab{}.
\newblock \showarticletitle{FlashRAG: A Modular Toolkit for Efficient Retrieval-Augmented Generation Research}.
\newblock \bibinfo{journal}{\emph{CoRR}}  \bibinfo{volume}{abs/2405.13576} (\bibinfo{year}{2024}).
\newblock
\showeprint[arXiv]{2405.13576}
\urldef\tempurl%
\url{https://arxiv.org/abs/2405.13576}
\showURL{%
\tempurl}


\bibitem[Kamalloo et~al\mbox{.}(2023)]%
        {kamalloo:2023}
\bibfield{author}{\bibinfo{person}{Ehsan Kamalloo}, \bibinfo{person}{Aref Jafari}, \bibinfo{person}{Xinyu Zhang}, \bibinfo{person}{Nandan Thakur}, {and} \bibinfo{person}{Jimmy Lin}.} \bibinfo{year}{2023}\natexlab{}.
\newblock \showarticletitle{{HAGRID:} {A} Human-LLM Collaborative Dataset for Generative Information-Seeking with Attribution}.
\newblock \bibinfo{journal}{\emph{CoRR}}  \bibinfo{volume}{abs/2307.16883} (\bibinfo{year}{2023}).
\newblock
\urldef\tempurl%
\url{https://doi.org/10.48550/ARXIV.2307.16883}
\showDOI{\tempurl}
\showeprint[arXiv]{2307.16883}


\bibitem[Khandelwal et~al\mbox{.}(2020)]%
        {khandelwal:2020}
\bibfield{author}{\bibinfo{person}{Urvashi Khandelwal}, \bibinfo{person}{Omer Levy}, \bibinfo{person}{Dan Jurafsky}, \bibinfo{person}{Luke Zettlemoyer}, {and} \bibinfo{person}{Mike Lewis}.} \bibinfo{year}{2020}\natexlab{}.
\newblock \showarticletitle{Generalization through Memorization: Nearest Neighbor Language Models}. In \bibinfo{booktitle}{\emph{8th International Conference on Learning Representations, {ICLR} 2020, Addis Ababa, Ethiopia, April 26-30, 2020}}. \bibinfo{publisher}{OpenReview.net}.
\newblock
\urldef\tempurl%
\url{https://openreview.net/forum?id=HklBjCEKvH}
\showURL{%
\tempurl}


\bibitem[Kulkarni et~al\mbox{.}(2020)]%
        {kulkarni:2020}
\bibfield{author}{\bibinfo{person}{Sayali Kulkarni}, \bibinfo{person}{Sheide Chammas}, \bibinfo{person}{Wan Zhu}, \bibinfo{person}{Fei Sha}, {and} \bibinfo{person}{Eugene Ie}.} \bibinfo{year}{2020}\natexlab{}.
\newblock \showarticletitle{AQuaMuSe: Automatically Generating Datasets for Query-Based Multi-Document Summarization}.
\newblock \bibinfo{journal}{\emph{CoRR}}  \bibinfo{volume}{abs/2010.12694} (\bibinfo{year}{2020}).
\newblock
\showeprint[arXiv]{2010.12694}
\urldef\tempurl%
\url{https://arxiv.org/abs/2010.12694}
\showURL{%
\tempurl}


\bibitem[Kwiatkowski et~al\mbox{.}(2019)]%
        {kwiatkowski:2019}
\bibfield{author}{\bibinfo{person}{Tom Kwiatkowski}, \bibinfo{person}{Jennimaria Palomaki}, \bibinfo{person}{Olivia Redfield}, \bibinfo{person}{Michael Collins}, \bibinfo{person}{Ankur~P. Parikh}, \bibinfo{person}{Chris Alberti}, \bibinfo{person}{Danielle Epstein}, \bibinfo{person}{Illia Polosukhin}, \bibinfo{person}{Jacob Devlin}, \bibinfo{person}{Kenton Lee}, \bibinfo{person}{Kristina Toutanova}, \bibinfo{person}{Llion Jones}, \bibinfo{person}{Matthew Kelcey}, \bibinfo{person}{Ming{-}Wei Chang}, \bibinfo{person}{Andrew~M. Dai}, \bibinfo{person}{Jakob Uszkoreit}, \bibinfo{person}{Quoc Le}, {and} \bibinfo{person}{Slav Petrov}.} \bibinfo{year}{2019}\natexlab{}.
\newblock \showarticletitle{Natural Questions: a Benchmark for Question Answering Research}.
\newblock \bibinfo{journal}{\emph{Trans. Assoc. Comput. Linguistics}}  \bibinfo{volume}{7} (\bibinfo{year}{2019}), \bibinfo{pages}{452--466}.
\newblock
\urldef\tempurl%
\url{https://doi.org/10.1162/TACL\_A\_00276}
\showDOI{\tempurl}


\bibitem[Lewis et~al\mbox{.}(2020)]%
        {lewis:2020}
\bibfield{author}{\bibinfo{person}{Patrick S.~H. Lewis}, \bibinfo{person}{Ethan Perez}, \bibinfo{person}{Aleksandra Piktus}, \bibinfo{person}{Fabio Petroni}, \bibinfo{person}{Vladimir Karpukhin}, \bibinfo{person}{Naman Goyal}, \bibinfo{person}{Heinrich K{\"{u}}ttler}, \bibinfo{person}{Mike Lewis}, \bibinfo{person}{Wen{-}tau Yih}, \bibinfo{person}{Tim Rockt{\"{a}}schel}, \bibinfo{person}{Sebastian Riedel}, {and} \bibinfo{person}{Douwe Kiela}.} \bibinfo{year}{2020}\natexlab{}.
\newblock \showarticletitle{Retrieval-Augmented Generation for Knowledge-Intensive {NLP} Tasks}. In \bibinfo{booktitle}{\emph{Advances in Neural Information Processing Systems 33: Annual Conference on Neural Information Processing Systems 2020, NeurIPS 2020, December 6-12, 2020, virtual}}, \bibfield{editor}{\bibinfo{person}{Hugo Larochelle}, \bibinfo{person}{Marc'Aurelio Ranzato}, \bibinfo{person}{Raia Hadsell}, \bibinfo{person}{Maria{-}Florina Balcan}, {and} \bibinfo{person}{Hsuan{-}Tien Lin}} (Eds.).
\newblock
\urldef\tempurl%
\url{https://proceedings.neurips.cc/paper/2020/hash/6b493230205f780e1bc26945df7481e5-Abstract.html}
\showURL{%
\tempurl}


\bibitem[Lin and Demner{-}Fushman(2006)]%
        {lin:2006}
\bibfield{author}{\bibinfo{person}{Jimmy Lin} {and} \bibinfo{person}{Dina Demner{-}Fushman}.} \bibinfo{year}{2006}\natexlab{}.
\newblock \showarticletitle{Methods for automatically evaluating answers to complex questions}.
\newblock \bibinfo{journal}{\emph{Inf. Retr.}} \bibinfo{volume}{9}, \bibinfo{number}{5} (\bibinfo{year}{2006}), \bibinfo{pages}{565--587}.
\newblock
\urldef\tempurl%
\url{https://doi.org/10.1007/S10791-006-9003-7}
\showDOI{\tempurl}


\bibitem[Lin et~al\mbox{.}(2021)]%
        {pyserini}
\bibfield{author}{\bibinfo{person}{Jimmy Lin}, \bibinfo{person}{Xueguang Ma}, \bibinfo{person}{Sheng-Chieh Lin}, \bibinfo{person}{Jheng-Hong Yang}, \bibinfo{person}{Ronak Pradeep}, {and} \bibinfo{person}{Rodrigo Nogueira}.} \bibinfo{year}{2021}\natexlab{}.
\newblock \showarticletitle{{Pyserini}: A {Python} Toolkit for Reproducible Information Retrieval Research with Sparse and Dense Representations}. In \bibinfo{booktitle}{\emph{Proceedings of the 44th Annual International ACM SIGIR Conference on Research and Development in Information Retrieval (SIGIR 2021)}}. \bibinfo{pages}{2356--2362}.
\newblock


\bibitem[Lin et~al\mbox{.}(2022)]%
        {lin:2022}
\bibfield{author}{\bibinfo{person}{Stephanie Lin}, \bibinfo{person}{Jacob Hilton}, {and} \bibinfo{person}{Owain Evans}.} \bibinfo{year}{2022}\natexlab{}.
\newblock \showarticletitle{TruthfulQA: Measuring How Models Mimic Human Falsehoods}. In \bibinfo{booktitle}{\emph{Proceedings of the 60th Annual Meeting of the Association for Computational Linguistics (Volume 1: Long Papers), {ACL} 2022, Dublin, Ireland, May 22-27, 2022}}, \bibfield{editor}{\bibinfo{person}{Smaranda Muresan}, \bibinfo{person}{Preslav Nakov}, {and} \bibinfo{person}{Aline Villavicencio}} (Eds.). \bibinfo{publisher}{Association for Computational Linguistics}, \bibinfo{pages}{3214--3252}.
\newblock
\urldef\tempurl%
\url{https://doi.org/10.18653/V1/2022.ACL-LONG.229}
\showDOI{\tempurl}


\bibitem[Liu(2022)]%
        {llamaindex}
\bibfield{author}{\bibinfo{person}{Jerry Liu}.} \bibinfo{year}{2022}\natexlab{}.
\newblock \bibinfo{booktitle}{\emph{LlamaIndex}}.
\newblock
\urldef\tempurl%
\url{https://www.llamaindex.ai/}
\showURL{%
\tempurl}


\bibitem[Liu et~al\mbox{.}(2024a)]%
        {liu:2024}
\bibfield{author}{\bibinfo{person}{Nelson~F. Liu}, \bibinfo{person}{Kevin Lin}, \bibinfo{person}{John Hewitt}, \bibinfo{person}{Ashwin Paranjape}, \bibinfo{person}{Michele Bevilacqua}, \bibinfo{person}{Fabio Petroni}, {and} \bibinfo{person}{Percy Liang}.} \bibinfo{year}{2024}\natexlab{a}.
\newblock \showarticletitle{{Lost in the Middle: How Language Models Use Long Contexts}}.
\newblock \bibinfo{journal}{\emph{Transactions of the Association for Computational Linguistics}}  \bibinfo{volume}{12} (\bibinfo{date}{02} \bibinfo{year}{2024}), \bibinfo{pages}{157--173}.
\newblock
\showISSN{2307-387X}
\urldef\tempurl%
\url{https://doi.org/10.1162/tacl_a_00638}
\showDOI{\tempurl}
\showeprint{https://direct.mit.edu/tacl/article-pdf/doi/10.1162/tacl\_a\_00638/2336043/tacl\_a\_00638.pdf}


\bibitem[Liu et~al\mbox{.}(2024b)]%
        {liu:2024b}
\bibfield{author}{\bibinfo{person}{Zihan Liu}, \bibinfo{person}{Wei Ping}, \bibinfo{person}{Rajarshi Roy}, \bibinfo{person}{Peng Xu}, \bibinfo{person}{Chankyu Lee}, \bibinfo{person}{Mohammad Shoeybi}, {and} \bibinfo{person}{Bryan Catanzaro}.} \bibinfo{year}{2024}\natexlab{b}.
\newblock \showarticletitle{ChatQA: Building {GPT-4} Level Conversational {QA} Models}.
\newblock \bibinfo{journal}{\emph{CoRR}}  \bibinfo{volume}{abs/2401.10225} (\bibinfo{year}{2024}).
\newblock
\urldef\tempurl%
\url{https://doi.org/10.48550/ARXIV.2401.10225}
\showDOI{\tempurl}
\showeprint[arXiv]{2401.10225}


\bibitem[Malaviya et~al\mbox{.}(2023)]%
        {malaviya:2023}
\bibfield{author}{\bibinfo{person}{Chaitanya Malaviya}, \bibinfo{person}{Subin Lee}, \bibinfo{person}{Sihao Chen}, \bibinfo{person}{Elizabeth Sieber}, \bibinfo{person}{Mark Yatskar}, {and} \bibinfo{person}{Dan Roth}.} \bibinfo{year}{2023}\natexlab{}.
\newblock \showarticletitle{ExpertQA: Expert-Curated Questions and Attributed Answers}.
\newblock \bibinfo{journal}{\emph{CoRR}}  \bibinfo{volume}{abs/2309.07852} (\bibinfo{year}{2023}).
\newblock
\urldef\tempurl%
\url{https://doi.org/10.48550/ARXIV.2309.07852}
\showDOI{\tempurl}
\showeprint[arXiv]{2309.07852}


\bibitem[Mayfield et~al\mbox{.}(2024)]%
        {mayfield:2024}
\bibfield{author}{\bibinfo{person}{James Mayfield}, \bibinfo{person}{Eugene Yang}, \bibinfo{person}{Dawn Lawrie}, \bibinfo{person}{Sean MacAvaney}, \bibinfo{person}{Paul McNamee}, \bibinfo{person}{Douglas~W. Oard}, \bibinfo{person}{Luca Soldaini}, \bibinfo{person}{Ian Soboroff}, \bibinfo{person}{Orion Weller}, \bibinfo{person}{Efsun Kayi}, \bibinfo{person}{Kate Sanders}, \bibinfo{person}{Marc Mason}, {and} \bibinfo{person}{Noah Hibbler}.} \bibinfo{year}{2024}\natexlab{}.
\newblock \showarticletitle{On the Evaluation of Machine-Generated Reports}. In \bibinfo{booktitle}{\emph{Proceedings of the 47th International ACM SIGIR Conference on Research and Development in Information Retrieval}}.
\newblock


\bibitem[Merrick et~al\mbox{.}(2024)]%
        {merrick:2024}
\bibfield{author}{\bibinfo{person}{Luke Merrick}, \bibinfo{person}{Danmei Xu}, \bibinfo{person}{Gaurav Nuti}, {and} \bibinfo{person}{Daniel Campos}.} \bibinfo{year}{2024}\natexlab{}.
\newblock \bibinfo{title}{Arctic-Embed: Scalable, Efficient, and Accurate Text Embedding Models}.
\newblock
\newblock
\showeprint[arxiv]{2405.05374}~[cs.CL]


\bibitem[Microsoft(2023)]%
        {bingsearch}
\bibfield{author}{\bibinfo{person}{Microsoft}.} \bibinfo{year}{2023}\natexlab{}.
\newblock \bibinfo{booktitle}{\emph{Reinventing search with a new {AI}-powered Microsoft Bing and Edge, your copilot for the web}}.
\newblock
\urldef\tempurl%
\url{https://blogs.microsoft.com/blog/2023/02/07/reinventing-search-with-a-new-ai-powered-microsoft-bing-and-edge-your-copilot-for-the-web/}
\showURL{%
\tempurl}


\bibitem[Muennighoff et~al\mbox{.}(2023)]%
        {muennighoff:2023}
\bibfield{author}{\bibinfo{person}{Niklas Muennighoff}, \bibinfo{person}{Nouamane Tazi}, \bibinfo{person}{Lo{\"{\i}}c Magne}, {and} \bibinfo{person}{Nils Reimers}.} \bibinfo{year}{2023}\natexlab{}.
\newblock \showarticletitle{{MTEB:} Massive Text Embedding Benchmark}. In \bibinfo{booktitle}{\emph{Proceedings of the 17th Conference of the European Chapter of the Association for Computational Linguistics, {EACL} 2023, Dubrovnik, Croatia, May 2-6, 2023}}, \bibfield{editor}{\bibinfo{person}{Andreas Vlachos} {and} \bibinfo{person}{Isabelle Augenstein}} (Eds.). \bibinfo{publisher}{Association for Computational Linguistics}, \bibinfo{pages}{2006--2029}.
\newblock
\urldef\tempurl%
\url{https://doi.org/10.18653/V1/2023.EACL-MAIN.148}
\showDOI{\tempurl}


\bibitem[OpenAI(2024)]%
        {openai_gpt4o}
\bibfield{author}{\bibinfo{person}{OpenAI}.} \bibinfo{year}{2024}\natexlab{}.
\newblock \bibinfo{booktitle}{\emph{Hello {GPT}-4o}}.
\newblock
\urldef\tempurl%
\url{https://openai.com/index/hello-gpt-4o/}
\showURL{%
\tempurl}


\bibitem[Overwijk et~al\mbox{.}(2022)]%
        {overwijk:2022}
\bibfield{author}{\bibinfo{person}{Arnold Overwijk}, \bibinfo{person}{Chenyan Xiong}, {and} \bibinfo{person}{Jamie Callan}.} \bibinfo{year}{2022}\natexlab{}.
\newblock \showarticletitle{ClueWeb22: 10 Billion Web Documents with Rich Information}. In \bibinfo{booktitle}{\emph{{SIGIR} '22: The 45th International {ACM} {SIGIR} Conference on Research and Development in Information Retrieval, Madrid, Spain, July 11 - 15, 2022}}, \bibfield{editor}{\bibinfo{person}{Enrique Amig{\'{o}}}, \bibinfo{person}{Pablo Castells}, \bibinfo{person}{Julio Gonzalo}, \bibinfo{person}{Ben Carterette}, \bibinfo{person}{J.~Shane Culpepper}, {and} \bibinfo{person}{Gabriella Kazai}} (Eds.). \bibinfo{publisher}{{ACM}}, \bibinfo{pages}{3360--3362}.
\newblock
\urldef\tempurl%
\url{https://doi.org/10.1145/3477495.3536321}
\showDOI{\tempurl}


\bibitem[Owoicho et~al\mbox{.}(2022)]%
        {owoicho:2022}
\bibfield{author}{\bibinfo{person}{Paul Owoicho}, \bibinfo{person}{Jeff Dalton}, \bibinfo{person}{Mohammad Aliannejadi}, \bibinfo{person}{Leif Azzopardi}, \bibinfo{person}{Johanne~R. Trippas}, {and} \bibinfo{person}{Svitlana Vakulenko}.} \bibinfo{year}{2022}\natexlab{}.
\newblock \showarticletitle{{TREC} CAsT 2022: Going Beyond User Ask and System Retrieve with Initiative and Response Generation}. In \bibinfo{booktitle}{\emph{Proceedings of the Thirty-First Text REtrieval Conference, {TREC} 2022, online, November 15-19, 2022}} \emph{(\bibinfo{series}{{NIST} Special Publication}, Vol.~\bibinfo{volume}{500-338})}, \bibfield{editor}{\bibinfo{person}{Ian Soboroff} {and} \bibinfo{person}{Angela Ellis}} (Eds.). \bibinfo{publisher}{National Institute of Standards and Technology {(NIST)}}.
\newblock
\urldef\tempurl%
\url{https://trec.nist.gov/pubs/trec31/papers/Overview\_cast.pdf}
\showURL{%
\tempurl}


\bibitem[Petroni et~al\mbox{.}(2021)]%
        {petroni:2021}
\bibfield{author}{\bibinfo{person}{Fabio Petroni}, \bibinfo{person}{Aleksandra Piktus}, \bibinfo{person}{Angela Fan}, \bibinfo{person}{Patrick S.~H. Lewis}, \bibinfo{person}{Majid Yazdani}, \bibinfo{person}{Nicola~De Cao}, \bibinfo{person}{James Thorne}, \bibinfo{person}{Yacine Jernite}, \bibinfo{person}{Vladimir Karpukhin}, \bibinfo{person}{Jean Maillard}, \bibinfo{person}{Vassilis Plachouras}, \bibinfo{person}{Tim Rockt{\"{a}}schel}, {and} \bibinfo{person}{Sebastian Riedel}.} \bibinfo{year}{2021}\natexlab{}.
\newblock \showarticletitle{{KILT:} a Benchmark for Knowledge Intensive Language Tasks}. In \bibinfo{booktitle}{\emph{Proceedings of the 2021 Conference of the North American Chapter of the Association for Computational Linguistics: Human Language Technologies, {NAACL-HLT} 2021, Online, June 6-11, 2021}}, \bibfield{editor}{\bibinfo{person}{Kristina Toutanova}, \bibinfo{person}{Anna Rumshisky}, \bibinfo{person}{Luke Zettlemoyer}, \bibinfo{person}{Dilek Hakkani{-}T{\"{u}}r}, \bibinfo{person}{Iz~Beltagy}, \bibinfo{person}{Steven Bethard}, \bibinfo{person}{Ryan Cotterell}, \bibinfo{person}{Tanmoy Chakraborty}, {and} \bibinfo{person}{Yichao Zhou}} (Eds.). \bibinfo{publisher}{Association for Computational Linguistics}, \bibinfo{pages}{2523--2544}.
\newblock
\urldef\tempurl%
\url{https://doi.org/10.18653/V1/2021.NAACL-MAIN.200}
\showDOI{\tempurl}


\bibitem[Pradeep et~al\mbox{.}(2021)]%
        {EMD}
\bibfield{author}{\bibinfo{person}{Ronak Pradeep}, \bibinfo{person}{Rodrigo Nogueira}, {and} \bibinfo{person}{Jimmy Lin}.} \bibinfo{year}{2021}\natexlab{}.
\newblock \showarticletitle{The Expando-Mono-Duo Design Pattern for Text Ranking with Pretrained Sequence-to-Sequence Models}.
\newblock \bibinfo{journal}{\emph{arXiv:2101.05667}} (\bibinfo{year}{2021}).
\newblock


\bibitem[Pradeep et~al\mbox{.}(2023a)]%
        {pradeep:2023}
\bibfield{author}{\bibinfo{person}{Ronak Pradeep}, \bibinfo{person}{Sahel Sharifymoghaddam}, {and} \bibinfo{person}{Jimmy Lin}.} \bibinfo{year}{2023}\natexlab{a}.
\newblock \showarticletitle{RankVicuna: Zero-Shot Listwise Document Reranking with Open-Source Large Language Models}.
\newblock \bibinfo{journal}{\emph{CoRR}}  \bibinfo{volume}{abs/2309.15088} (\bibinfo{year}{2023}).
\newblock
\urldef\tempurl%
\url{https://doi.org/10.48550/ARXIV.2309.15088}
\showDOI{\tempurl}
\showeprint[arXiv]{2309.15088}


\bibitem[Pradeep et~al\mbox{.}(2023b)]%
        {pradeep:2023b}
\bibfield{author}{\bibinfo{person}{Ronak Pradeep}, \bibinfo{person}{Sahel Sharifymoghaddam}, {and} \bibinfo{person}{Jimmy Lin}.} \bibinfo{year}{2023}\natexlab{b}.
\newblock \showarticletitle{RankZephyr: Effective and Robust Zero-Shot Listwise Reranking is a Breeze!}
\newblock \bibinfo{journal}{\emph{CoRR}}  \bibinfo{volume}{abs/2312.02724} (\bibinfo{year}{2023}).
\newblock
\urldef\tempurl%
\url{https://doi.org/10.48550/ARXIV.2312.02724}
\showDOI{\tempurl}
\showeprint[arXiv]{2312.02724}


\bibitem[Raina and Gales(2024)]%
        {raina:2024}
\bibfield{author}{\bibinfo{person}{Vatsal Raina} {and} \bibinfo{person}{Mark Gales}.} \bibinfo{year}{2024}\natexlab{}.
\newblock \bibinfo{title}{Question-Based Retrieval using Atomic Units for Enterprise RAG}.
\newblock
\newblock
\showeprint[arxiv]{2405.12363}~[cs.CL]


\bibitem[Robertson and Zaragoza(2009)]%
        {robertson:2009}
\bibfield{author}{\bibinfo{person}{Stephen~E. Robertson} {and} \bibinfo{person}{Hugo Zaragoza}.} \bibinfo{year}{2009}\natexlab{}.
\newblock \showarticletitle{The Probabilistic Relevance Framework: {BM25} and Beyond}.
\newblock \bibinfo{journal}{\emph{Found. Trends Inf. Retr.}} \bibinfo{volume}{3}, \bibinfo{number}{4} (\bibinfo{year}{2009}), \bibinfo{pages}{333--389}.
\newblock
\urldef\tempurl%
\url{https://doi.org/10.1561/1500000019}
\showDOI{\tempurl}


\bibitem[Rosenthal et~al\mbox{.}(2024)]%
        {rosenthal:2024}
\bibfield{author}{\bibinfo{person}{Sara Rosenthal}, \bibinfo{person}{Avirup Sil}, \bibinfo{person}{Radu Florian}, {and} \bibinfo{person}{Salim Roukos}.} \bibinfo{year}{2024}\natexlab{}.
\newblock \showarticletitle{{CLAPNQ:} Cohesive Long-form Answers from Passages in Natural Questions for {RAG} systems}.
\newblock \bibinfo{journal}{\emph{CoRR}}  \bibinfo{volume}{abs/2404.02103} (\bibinfo{year}{2024}).
\newblock
\urldef\tempurl%
\url{https://doi.org/10.48550/ARXIV.2404.02103}
\showDOI{\tempurl}
\showeprint[arXiv]{2404.02103}


\bibitem[Rosset et~al\mbox{.}(2024)]%
        {rosset:2024}
\bibfield{author}{\bibinfo{person}{Corby Rosset}, \bibinfo{person}{Ho{-}Lam Chung}, \bibinfo{person}{Guanghui Qin}, \bibinfo{person}{Ethan~C. Chau}, \bibinfo{person}{Zhuo Feng}, \bibinfo{person}{Ahmed Awadallah}, \bibinfo{person}{Jennifer Neville}, {and} \bibinfo{person}{Nikhil Rao}.} \bibinfo{year}{2024}\natexlab{}.
\newblock \showarticletitle{Researchy Questions: {A} Dataset of Multi-Perspective, Decompositional Questions for {LLM} Web Agents}.
\newblock \bibinfo{journal}{\emph{CoRR}}  \bibinfo{volume}{abs/2402.17896} (\bibinfo{year}{2024}).
\newblock
\urldef\tempurl%
\url{https://doi.org/10.48550/ARXIV.2402.17896}
\showDOI{\tempurl}
\showeprint[arXiv]{2402.17896}


\bibitem[Stelmakh et~al\mbox{.}(2022)]%
        {stelmark:2022}
\bibfield{author}{\bibinfo{person}{Ivan Stelmakh}, \bibinfo{person}{Yi Luan}, \bibinfo{person}{Bhuwan Dhingra}, {and} \bibinfo{person}{Ming{-}Wei Chang}.} \bibinfo{year}{2022}\natexlab{}.
\newblock \showarticletitle{{ASQA:} Factoid Questions Meet Long-Form Answers}. In \bibinfo{booktitle}{\emph{Proceedings of the 2022 Conference on Empirical Methods in Natural Language Processing, {EMNLP} 2022, Abu Dhabi, United Arab Emirates, December 7-11, 2022}}, \bibfield{editor}{\bibinfo{person}{Yoav Goldberg}, \bibinfo{person}{Zornitsa Kozareva}, {and} \bibinfo{person}{Yue Zhang}} (Eds.). \bibinfo{publisher}{Association for Computational Linguistics}, \bibinfo{pages}{8273--8288}.
\newblock
\urldef\tempurl%
\url{https://doi.org/10.18653/V1/2022.EMNLP-MAIN.566}
\showDOI{\tempurl}


\bibitem[Sun et~al\mbox{.}(2023)]%
        {rankgpt}
\bibfield{author}{\bibinfo{person}{Weiwei Sun}, \bibinfo{person}{Lingyong Yan}, \bibinfo{person}{Xinyu Ma}, \bibinfo{person}{Pengjie Ren}, \bibinfo{person}{Dawei Yin}, {and} \bibinfo{person}{Zhaochun Ren}.} \bibinfo{year}{2023}\natexlab{}.
\newblock \showarticletitle{Is ChatGPT Good at Search? Investigating Large Language Models as Re-Ranking Agent}.
\newblock \bibinfo{journal}{\emph{arXiv:2304.09542}} (\bibinfo{year}{2023}).
\newblock


\bibitem[Thakur et~al\mbox{.}(2021)]%
        {thakur:2021}
\bibfield{author}{\bibinfo{person}{Nandan Thakur}, \bibinfo{person}{Nils Reimers}, \bibinfo{person}{Andreas R{\"{u}}ckl{\'{e}}}, \bibinfo{person}{Abhishek Srivastava}, {and} \bibinfo{person}{Iryna Gurevych}.} \bibinfo{year}{2021}\natexlab{}.
\newblock \showarticletitle{{BEIR:} {A} Heterogeneous Benchmark for Zero-shot Evaluation of Information Retrieval Models}. In \bibinfo{booktitle}{\emph{Proceedings of the Neural Information Processing Systems Track on Datasets and Benchmarks 1, NeurIPS Datasets and Benchmarks 2021, December 2021, virtual}}, \bibfield{editor}{\bibinfo{person}{Joaquin Vanschoren} {and} \bibinfo{person}{Sai{-}Kit Yeung}} (Eds.).
\newblock
\urldef\tempurl%
\url{https://datasets-benchmarks-proceedings.neurips.cc/paper/2021/hash/65b9eea6e1cc6bb9f0cd2a47751a186f-Abstract-round2.html}
\showURL{%
\tempurl}


\bibitem[Xiong et~al\mbox{.}(2024)]%
        {xiong:2024}
\bibfield{author}{\bibinfo{person}{Guangzhi Xiong}, \bibinfo{person}{Qiao Jin}, \bibinfo{person}{Zhiyong Lu}, {and} \bibinfo{person}{Aidong Zhang}.} \bibinfo{year}{2024}\natexlab{}.
\newblock \showarticletitle{Benchmarking Retrieval-Augmented Generation for Medicine}.
\newblock \bibinfo{journal}{\emph{CoRR}}  \bibinfo{volume}{abs/2402.13178} (\bibinfo{year}{2024}).
\newblock
\urldef\tempurl%
\url{https://doi.org/10.48550/ARXIV.2402.13178}
\showDOI{\tempurl}
\showeprint[arXiv]{2402.13178}


\bibitem[Yan et~al\mbox{.}(2024)]%
        {yan:2024}
\bibfield{author}{\bibinfo{person}{Shi{-}Qi Yan}, \bibinfo{person}{Jia{-}Chen Gu}, \bibinfo{person}{Yun Zhu}, {and} \bibinfo{person}{Zhen{-}Hua Ling}.} \bibinfo{year}{2024}\natexlab{}.
\newblock \showarticletitle{Corrective Retrieval Augmented Generation}.
\newblock \bibinfo{journal}{\emph{CoRR}}  \bibinfo{volume}{abs/2401.15884} (\bibinfo{year}{2024}).
\newblock
\urldef\tempurl%
\url{https://doi.org/10.48550/ARXIV.2401.15884}
\showDOI{\tempurl}
\showeprint[arXiv]{2401.15884}


\bibitem[Yang et~al\mbox{.}(2017)]%
        {anserini}
\bibfield{author}{\bibinfo{person}{Peilin Yang}, \bibinfo{person}{Hui Fang}, {and} \bibinfo{person}{Jimmy Lin}.} \bibinfo{year}{2017}\natexlab{}.
\newblock \showarticletitle{Anserini: Enabling the Use of Lucene for Information Retrieval Research}. In \bibinfo{booktitle}{\emph{International Conference on Research and Development in Information Retrieval (SIGIR)}}.
\newblock
\urldef\tempurl%
\url{https://doi.org/10.1145/3077136.3080721}
\showDOI{\tempurl}


\bibitem[Zheng et~al\mbox{.}(2023)]%
        {zheng:2024}
\bibfield{author}{\bibinfo{person}{Lianmin Zheng}, \bibinfo{person}{Wei{-}Lin Chiang}, \bibinfo{person}{Ying Sheng}, \bibinfo{person}{Siyuan Zhuang}, \bibinfo{person}{Zhanghao Wu}, \bibinfo{person}{Yonghao Zhuang}, \bibinfo{person}{Zi Lin}, \bibinfo{person}{Zhuohan Li}, \bibinfo{person}{Dacheng Li}, \bibinfo{person}{Eric~P. Xing}, \bibinfo{person}{Hao Zhang}, \bibinfo{person}{Joseph~E. Gonzalez}, {and} \bibinfo{person}{Ion Stoica}.} \bibinfo{year}{2023}\natexlab{}.
\newblock \showarticletitle{Judging LLM-as-a-Judge with MT-Bench and Chatbot Arena}. In \bibinfo{booktitle}{\emph{Advances in Neural Information Processing Systems 36: Annual Conference on Neural Information Processing Systems 2023, NeurIPS 2023, New Orleans, LA, USA, December 10 - 16, 2023}}, \bibfield{editor}{\bibinfo{person}{Alice Oh}, \bibinfo{person}{Tristan Naumann}, \bibinfo{person}{Amir Globerson}, \bibinfo{person}{Kate Saenko}, \bibinfo{person}{Moritz Hardt}, {and} \bibinfo{person}{Sergey Levine}} (Eds.).
\newblock
\urldef\tempurl%
\url{http://papers.nips.cc/paper\_files/paper/2023/hash/91f18a1287b398d378ef22505bf41832-Abstract-Datasets\_and\_Benchmarks.html}
\showURL{%
\tempurl}


\end{thebibliography}

%%
%% If your work has an appendix, this is the place to put it.
\clearpage
\appendix

\section{TREC-RAGgy 2024: Additional Details}\label{sec:trec-raggy-appendix}

We manually classify each available topic in TREC Deep Learning Tracks 2021-2023 \cite{craswell:2021, craswell:2022, chiang:2024} into one of the seven different topic categories. We manually labeled each topic following the guidelines\footnote{Guidelines have been inspired from the \href{https://www.aicrowd.com/challenges/meta-comprehensive-rag-benchmark-kdd-cup-2024}{2024 Meta Comprehensive RAG benchmark}.} mentioned below:
\begin{itemize}
    \item \emph{Simple}: topic asking for information about a simple fact, e.g., ``how to emulsion a house?''
    \item \emph{Simple with condition}: topic asking for information about a topic with an imposed condition, e.g., ``how to cook thinly sliced home fries?''
    \item \emph{Set}: a topic containing multiple short entities in the answer, e.g., ``what themes are in action movies?''
    \item \emph{Aggregation}: a topic that requires aggregation of multiple retrieved segments, e.g., ``how to put together a scuba regulator?''
    \item \emph{Comparison}: a topic that requires comparison of the retrieved segments, e.g. ``does light intensity or concentration of carbon dioxide have a higher rate of photosynthesis?''
    \item \emph{Multi-hop}: a topic that requires to chain multiple information from different retrieved segments, e.g., ``the population of kings grant fayetteville prior to liberty hills?''
    \item \emph{False premise}: a topic that has a false preposition or assumption, e.g., ``Do larger lobsters become tougher when cooked?''
\end{itemize}

\section{TREC-Researchy 2024: Additional Details}\label{sec:trec-researchy-appendix}
Note that for Researchy Questions~\cite{rosset:2024}, the following eight intrinsic attributes were measured by GPT-4 on a scale of 0-10:
\begin{itemize}
    \item \emph{Ambiguity:} Checks if the question’s intent is moderately ambiguous, suggesting multiple interpretations.
    \item \emph{Incompleteness:} Checks if the question is difficult to answer due to missing crucial context or details.
    \item \emph{Assumptive:} Checks if the question has some built-in assumptions that may influence the answer.
    \item \emph{Multi-faceted:} Checks if the question requires considering multiple perspectives to provide a comprehensive answer.
    \item \emph{Knowledge-intensive:} Checks if the question demands specialized knowledge and extensive research to answer thoroughly.
    \item \emph{Subjective:} Measures if the question contains some level of subjectivity, with potential for varying opinions.
    \item \emph{Reasoning-intensive:} Checks if the question requires significant reasoning and synthesis of information to answer.
    \item \emph{Harmful:} Checks to what extent the question is harmful or inappropriate.
\end{itemize}
It is worth noting that all the questions provided scored 0 in harmfulness and a tiny fraction scored highly on ambiguity.
We used a score of $5$ as the threshold to label the query for that intrinsic attribute.

\section{\ragnarok System Arena}
\label{app:system_arena}

\begin{figure*}[htb]
    \centering
    \includegraphics[width=0.67\textwidth]
    {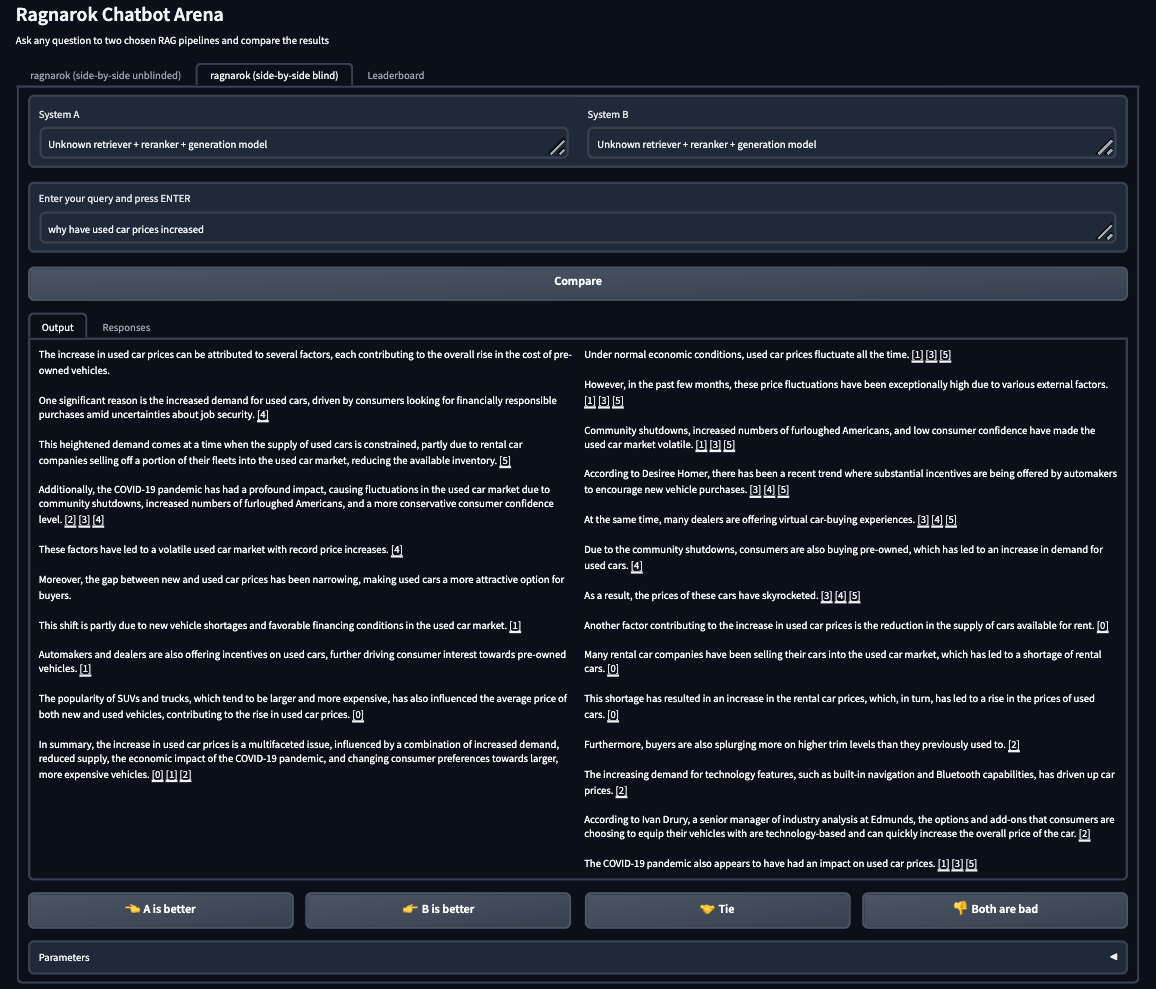}%
    \caption{
     WebUI (dark mode) showcasing the \ragnarok system arena for the user query on ``why have used car prices increased'' from TREC-2024 Researchy with two different blinded pipelines.
       The output tab displays the answers in human-readable form.
       }
    \label{fig:ragnarok-battle-blind}
\end{figure*}

\autoref{fig:ragnarok-battle-blind} showcases the \ragnarok WebUI (dark mode) and the user query, “why have used car prices increased”, from TREC-2024 Researchy issued to two blinded systems.
This blind setup enables fair leaderboards, especially when incentives to game leaderboards are huge in this competitive proprietary LLM space. 
The output displays the answers in human-readable form, allowing users to assess the quality of responses without bias.

\begin{figure*}[htb]
    \centering
    \includegraphics[width=0.67\textwidth]
    {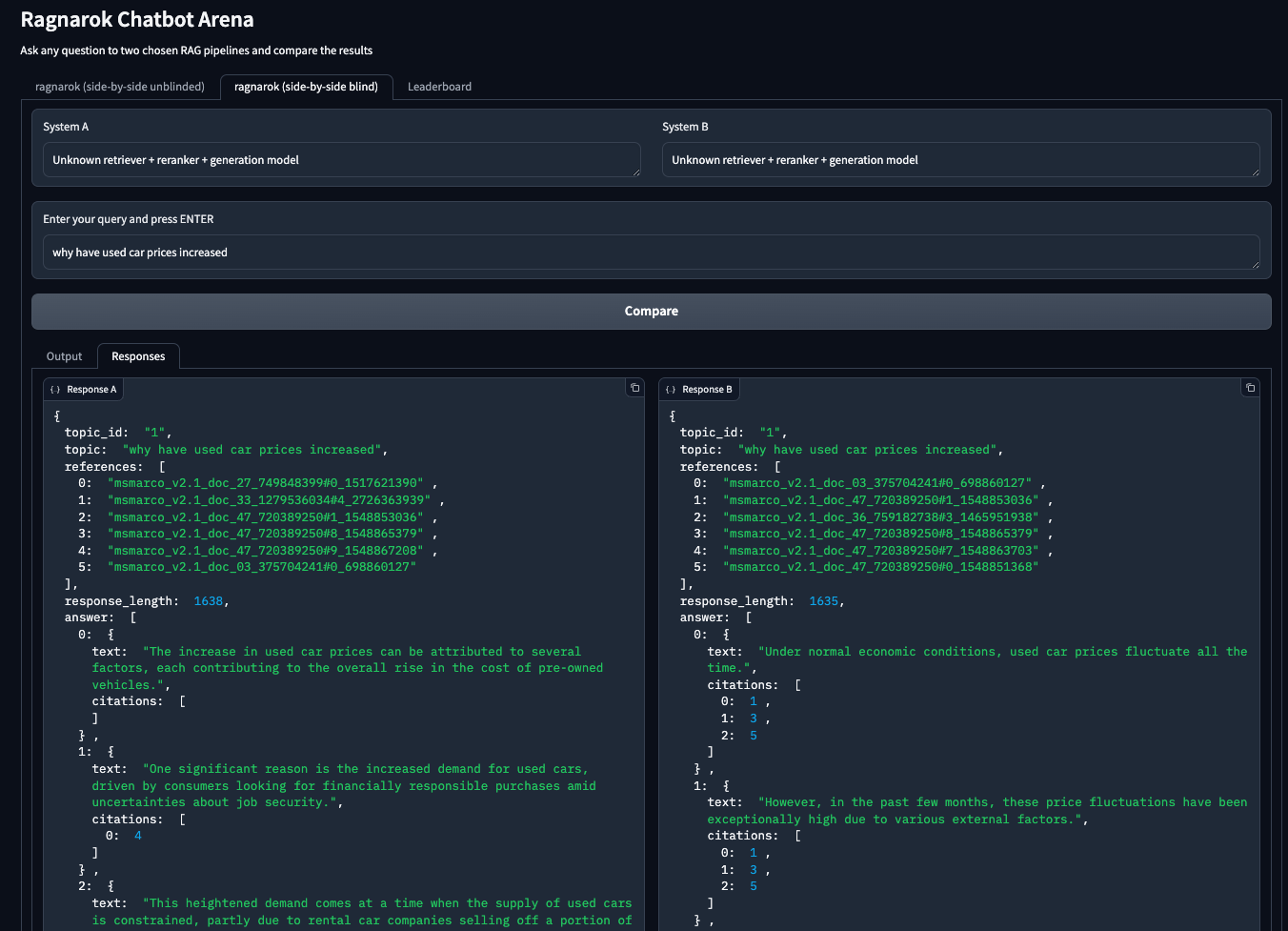}%
    \caption{
    The responses tab for the example in \autoref{fig:ragnarok-battle-blind}. Note that the responses tab reformats the final answers into the JSON format expected by the I/O definitions of the TREC 2024 RAG Track.
       }
    \label{fig:ragnarok-battle-blind-response}
\end{figure*}

\autoref{fig:ragnarok-battle-blind-response} demonstrates the responses tab for the example in \autoref{fig:ragnarok-battle-blind}.
 The responses tab reformats the final answers into the JSON output expected by the I/O definitions of the TREC 2024 RAG Track. 
This feature is particularly useful for developers and researchers who need to ensure that their systems’ outputs conform to specific standards and formats required by evaluation frameworks.

By incorporating both human-readable and JSON-formatted outputs, \ragnarok provides a comprehensive evaluation platform that caters to a wide range of needs in the research and development community. 
The ability to toggle between different views and formats ensures that users can efficiently analyze and interpret the effectiveness of various RAG systems.

\begin{figure}[t!]
\begin{mdframed}[backgroundcolor=gray!5]
    \small
    \texttt{System: This is a chat between a user and an artificial intelligence assistant. The assistant gives helpful, detailed, and polite answers to the user's questions based on the context. The assistant should also indicate when the answer cannot be found in the context. \\\\
    INSTRUCTION: Please give a complete answer to the question. Cite each context document that supports your answer within brackets [] using the IEEE format. \\\\
    QUESTION: \{query\} \\\\
    CONTEXTS:\\
    {[1]} \{Passage title\}: {\{Passage text\}}\\
    {[2]} \{Passage title\}: {\{Passage text\}}\\
    ... \\
    {[20]} \{Passage title\}: {\{Passage text\}}\\\\
    INSTRUCTION: Please give a complete answer to the question. Cite each context document that supports your answer within brackets [] using the IEEE format.
}
\end{mdframed}
\caption{ChatQA prompt template~\cite{liu:2024b} used for RAG generation with in-text citations with GPT-4o in our \ragnarok framework.}
\label{fig:prompt}
\end{figure}

\begin{figure*}[htb]
    \centering
    \includegraphics[clip, width=0.67\textwidth]{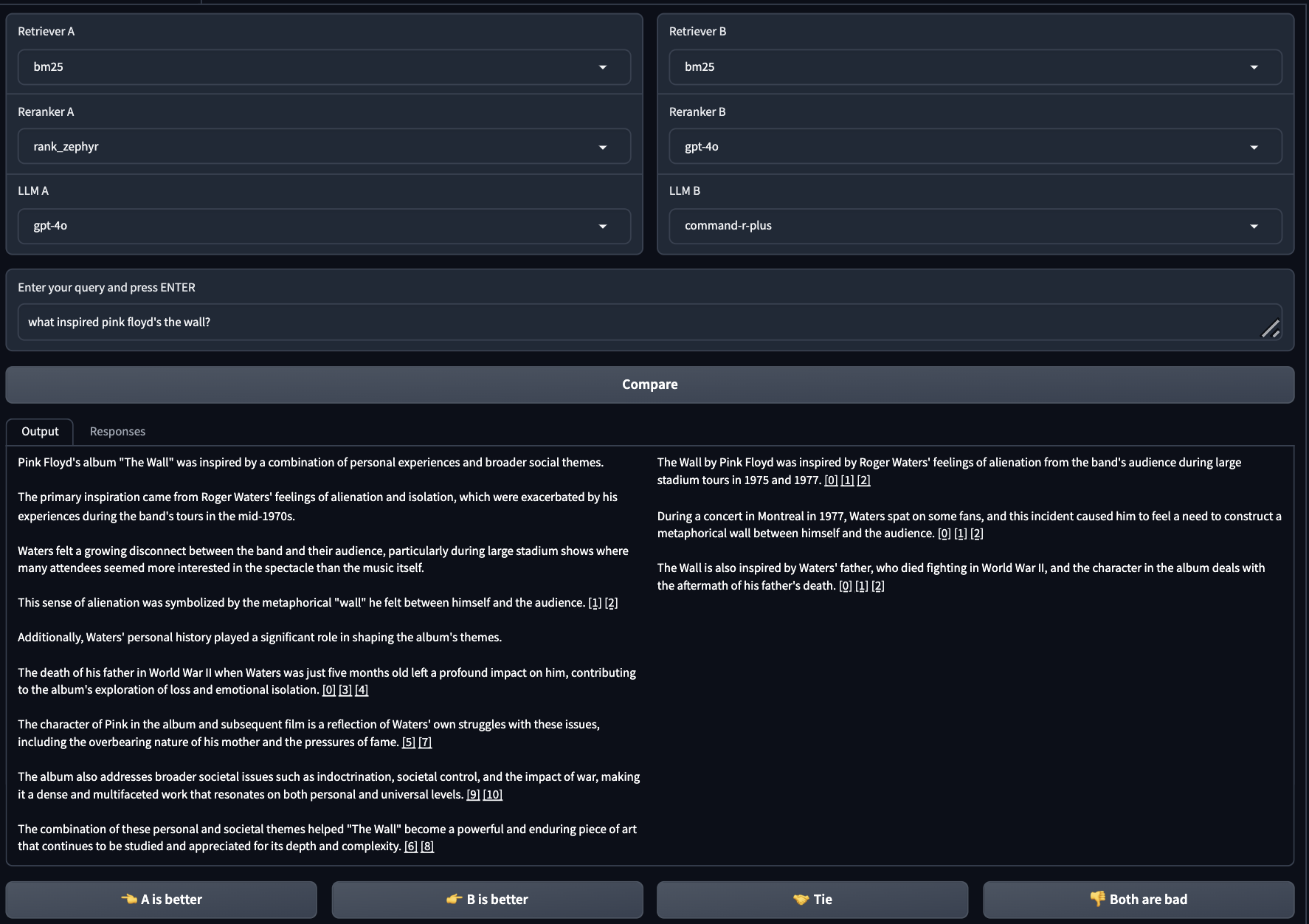}%
    \vspace{-1em}
    \caption{%
       WebUI showcasing the \ragnarok System Arena and the user query, ``what inspired pink floyd's the wall?'', with answers from two pipelines side-by-side comparing GPT-4o answer (left) and Command R+ answer (right).
       }%
       \vspace{2em}
    \label{fig:ragnarok-battle}%
\end{figure*}

%% add tables in the end
\begin{table*}[t]
\setlength{\tabcolsep}{3pt}
\caption{An example of the first segment of two near-duplicate documents present in the MS MARCO V2 segment collection. During the deduplication procedure, the segments of one of the documents is kept in the MS MARCO V2.1 segment collection (msmarco\_doc\_00\_995170174\#0), whereas the other segment is discarded as a duplicate (msmarco\_doc\_00\_995171191\#0).}
    \renewcommand{\arraystretch}{1}
    \small
    {\begin{tabular}{p{2cm} p{3.2cm} p{2cm} p{10cm}}
    \toprule
    \textbf{Segment ID} & \textbf{URL} & \textbf{Title} & \textbf{Segment} \\ \midrule
    \multicolumn{4}{l}{\textbf{MS MARCO V2.1 segment (discarded) }} \\ \midrule
    msmarco\_doc\_00 \_995171191\#0 & \href{http://center.serve.org/TT/fp\_tips.html}{http://center.serve.org/TT /fp\_tips.html} & SERVE Center Resources & SERVE Center Get to know us About Us We believe that educational improvement requires a partnership with our clients, and we focus on doing work that is important and directly relevant to both policymakers and practitioners. Services Program evaluation, capacity building, technical assistance, strategic planning, and much more. Customized services designed to meet the specific needs of our partnering organizations. Projects During its history, SERVE has been awarded over \$200 million in contracts and grants and has successfully managed 14 major awards, including multiple contracts with the US Department of Education. Resources Forging connections with a variety of individuals and organizations, the SERVE Center offers an array of resources developed to inform and aide both the policymaker and the practioner. \\  \midrule
    \multicolumn{4}{l}{\textbf{MS MARCO V2.1 segment (kept)}} \\ \midrule
    msmarco\_doc\_00 \_995170174\#0 & \href{http://center.serve.org/TT/}{http://center.serve.org/TT/} & SERVE Center Resources & SERVE Center Get to know us About Us We believe that educational improvement requires a partnership with our clients, and we focus on doing work that is important and directly relevant to both policymakers and practitioners. Services Program evaluation, capacity building, technical assistance, strategic planning, and much more. Customized services designed to meet the specific needs of our partnering organizations. Projects During its history, SERVE has been awarded over \$200 million in contracts and grants and has successfully managed 14 major awards, including multiple contracts with the US Department of Education. Resources Forging connections with a variety of individuals and organizations, the SERVE Center offers an array of resources developed to inform and aide both the policymaker and the practioner. \\  \midrule
\end{tabular}}\label{tab:duplicate-msmarco}
\end{table*}

\begin{table*}[t]
\setlength{\tabcolsep}{3pt}
\caption{An end-to-end RAG example for a randomly sampled topic (topic ID: 2027497) in the TREC-RAGgy 2024 collection: ``how often should you take your toddler to the potty when potty training?''}

\renewcommand{\arraystretch}{1}
\small
{\begin{tabular}{p{4.5cm} p{4cm} p{9.5cm}}
\toprule
% \multicolumn{3}{l}{User Topic: \textit{how often should you take your toddler to the potty when potty training?}} \\ \midrule
\multicolumn{3}{l}{\textbf{1. Top-10 retrieved segments from the (R) retrieval stage (ordered from top to bottom in terms of relevancy): }} \\  \midrule
\textbf{Segment ID} & \textbf{Title} & \textbf{Segment} \\ \noalign{\vskip 1ex}\hdashline\noalign{\vskip 1ex} 
(1) msmarco\_v2.1\_doc\_51\_766815931 \#2\_1606878413 & How Often Should I Take My Toddler To The Potty? There Are Some Guidelines To Follow & Self-esteem can be fragile at this time, so it's important to toilet train gently, letting your child lead the way. Additionally, it's unwise to begin potty training unless your child is truly ready. Take this potty training readiness quiz featured in Parents before you begin. Elizabeth Pantley [ ... ] and you still must be on the lookout for signs your toddler has to go ahead of schedule. \\  \noalign{\vskip 0.5ex}\noalign{\vskip 0.5ex} 
(2) msmarco\_v2.1\_doc\_51\_766815931 \#6\_1606884302 & How Often Should I Take My Toddler To The Potty? There Are Some Guidelines To Follow & Davis and Keyser stressed the importance of keeping your own emotions in check throughout the toilet training process. The bottom line: when it comes to potty time, there's no magic number, [ ... ] You'll just have that much longer to plan a good one. \\  \noalign{\vskip 0.5ex}\noalign{\vskip 0.5ex} 
(3) msmarco\_v2.1\_doc\_08\_935420812 \#8\_1683502976 & Q\&A: Know Your Child’s Level of Readiness for Potty Training | Parents & Start by sitting on the toilet the first thing in the morning after taking off the overnight diaper, make it fun—sing songs, read a book, drink some juice, and see what happens. [ ... ] — Dr. Carrie M. Brown My 3-year-old is adamant about not using the potty. How can I motivate her? \\ \noalign{\vskip 0.5ex}\noalign{\vskip 0.5ex} 
(4) msmarco\_v2.1\_doc\_08\_935420812 \#0\_1683481876 & Q\&A: Know Your Child’s Level of Readiness for Potty Training | Parents & Q\&A: Know Your Child’s Level of Readiness for Potty Training | Parents Home Toddlers \& Preschoolers Potty Training Potty Training Tips Q\&A: [ ... ] Have your child sit on the potty soon after she's finished breakfast and again after dinner. \\ \noalign{\vskip 0.5ex}\noalign{\vskip 0.5ex} 
(5) msmarco\_v2.1\_doc\_28\_472446307 \#23\_1012991039 & Potty Trained Toddler Having Accidents on Purpose? & But don’t feel disappointed if most of these tries end up with no pee in the potty—you wouldn’t be able to pee [ ... ] Get more tips on how to ease your child’s potty training poop anxiety. \\ \noalign{\vskip 0.5ex}\noalign{\vskip 0.5ex} 
(6) msmarco\_v2.1\_doc\_28\_472446307 \#22\_1012988885 & Potty Trained Toddler Having Accidents on Purpose? & Get tips on what to do when your 4 year old won’t poop on potty. [ ... ] Routines give him the predictability he needs so he knows exactly what to do and when. You might use the potty after waking up, before leaving the house or after eating meals. \\ \noalign{\vskip 0.5ex}\noalign{\vskip 0.5ex} 
(7) msmarco\_v2.1\_doc\_57\_1222573163 \#16\_2241797192 & Tips on Starting Potty Training: How to Potty Train Your Toddler & Teach him to check for dryness. This offers him an added sense of control. If he's dry, give him a pat on the back (or a big hug), but don't criticize him if he's wet. [ ... ] If you overreact, you might discourage your toddler's future attempts. \\ \noalign{\vskip 0.5ex}\noalign{\vskip 0.5ex} 
(8) msmarco\_v2.1\_doc\_57\_1222573163 \#15\_2241795642 & Tips on Starting Potty Training: How to Potty Train Your Toddler & Even if you're too late and he's already done the deed, have him sit on the potty anyway to reinforce the connection. Keep her motivated. [ ... ] Even the most enthusiastic toddler can take several weeks to master potty training proficiency — often with as many steps backward as forward. \\ \noalign{\vskip 0.5ex}\noalign{\vskip 0.5ex} 
(9) msmarco\_v2.1\_doc\_51\_766815931 \#5\_1606882767 & How Often Should I Take My Toddler To The Potty? There Are Some Guidelines To Follow & According to Becoming the Parent You Want To Be by Laura Davis and Janis Keyser, it's common for children to go back and forth — somedays preferring diapers, and other days selecting to use the toilet. [ ... ] You can only take your child to the bathroom too much if it begins to feel stressful or punitive. \\ \noalign{\vskip 0.5ex}\noalign{\vskip 0.5ex} 
(10) msmarco\_v2.1\_doc\_51\_766815931 \#0\_1606874600 & How Often Should I Take My Toddler To The Potty? There Are Some Guidelines To Follow & How Often Should I Take My Toddler To The Potty? There Are Some Guidelines To Follow Life Ashley Batz/Romper How Often Should I Take My Toddler To The Potty? There Are Some Guidelines To Follow by Kelly Mullen-McWilliams July 22, 2017 I don't know about you, but I'm personally super pumped for that perfect day  [ ... ] he article noted that toilet training is also one of the most difficult milestones for children and their parents, and that it can become highly emotional. \\ \midrule
\multicolumn{3}{l}{\textbf{2. Results from the (AG) augmented generation stage using top-10 retrieved segments: }} \\ \midrule
 \textbf{Model} & \multicolumn{2}{c}{\textbf{RAG Answer}} \\ \noalign{\vskip 1ex}\hdashline\noalign{\vskip 1ex} 
 Command R+ & \multicolumn{2}{p{14cm}}{[\{'text': 'There is no magic number for how often you should take your toddler to the potty when potty training.', 'citations': [1, 8]\}, \{'text': 'However, it is recommended to take them frequently, especially after waking up, before leaving the house, after eating meals, or before bath time.', 'citations': [0, 1, 2, 3, 4, 5, 8]\}, \{'text': 'You can also use a timer and set it to every 30 minutes to an hour for a potty break.', 'citations': [5]\}, \{'text': 'It is important to be patient and gentle throughout the process, as it can be emotionally challenging for both the child and the parent.', 'citations': [0, 1, 3, 6, 7, 8, 9]\}]} \\ \noalign{\vskip 1ex}\noalign{\vskip 1ex} 
 GPT-4o & \multicolumn{2}{p{14cm}}{[\{'text': 'When potty training your toddler, it is recommended to take them to the potty frequently to increase the chances of success.', 'citations': []\}, \{'text': 'According to Elizabeth Pantley, author of "The No-Cry Potty Training Solution," most toddlers pee four to eight times per day and have one or two bowel movements daily.', 'citations': []\}, \{'text': 'She suggests setting up a potty routine, such as taking your toddler to the potty first thing in the morning, after eating, and before activities like riding in the car or going to sleep.', 'citations': [0]\}, \{'text': 'Additionally, Davis and Keyser emphasize the importance of following a schedule, such as morning pee and before nap time, while also being attentive to signs that your toddler needs to go.', 'citations': []\}, \{'text': 'They caution against making the process stressful or punitive, as negative emotions can hinder progress.', 'citations': [1]\}, \{'text': 'Dr. Carrie M. Brown recommends planning bathroom trips every 90-120 minutes during the day and remaining calm and consistent, even if accidents occur.', 'citations': []\}, \{'text': 'This approach helps toddlers learn the routine and increases the likelihood of successful potty use.', 'citations': [2]\}]} \\ \bottomrule
\end{tabular}}\label{tab:e2e-trec-raggy}
\end{table*}

\end{document}